\documentclass[12pt,preprint]{aastex}

\newcommand{\units}[1]{\,\textrm{#1}}
\newcommand{\lap}{\mathrel{\vcenter{\hbox{\ooalign{\raise 4.75pt
\hbox{$<$}\crcr $\sim$}}}}}
\newcommand{\gap}{\mathrel{\vcenter{\hbox{\ooalign{\raise 4.75pt
\hbox{$>$}\crcr $\sim$}}}}}

\def\Msun{\hbox{$M_\odot$}}

\def\Ystar{\hbox{$Y_*$}}
\def\alfa{\hbox{$\alpha $}}
\def\alp2{\hbox{$\alpha^2$}}

\def\alpTH{\hbox{$\alpha_{\sf Th}$}}

\def\jHalp4pi{\hbox{$4 \pi j_{{\rm H} \alpha}$}}

\def\xTH{\hbox{$x_{\sf Th}$}}
\def\xinner{\hbox{$x_{\rm inner}$}}

\def\xouter{\hbox{$x_{\rm outer}$}}
\def\xalfven{\hbox{$x_{\rm Alf}$}}

\def\vphiXout{\hbox{$v_{\phi}(x_{\rm outer})$}}
\def\xpeak{\hbox{$x_{\rm peak}$}}
\def\dmin{\hbox{$d_{\rm inner}$}}
\def\dmax{\hbox{$d_{\rm outer}$}}
\def\dpeak{\hbox{$d_{\rm peak}$}}
\def\Alfven{\hbox{Alfv\'{e}n}}
\def\Bvec{\hbox{$\bf B $}}

\def\Vvec{\hbox{$\bf v$}}
\def\VM{\hbox{$V_M$}}

\def\OmegZ{\hbox{$\Omega_{\circ}$}}

\def\lami{\hbox{$l$}}

\def\SK{\hbox{$\cal S$}}
\def\SZ{\hbox{$\cal S_{\circ}$}}
\def\FM{\hbox{${\cal F}_{\sf M}$}}
\def\FMZ{\hbox{${\cal F}_{{\sf M},{\circ}}$}}
\def\BZ{\hbox{$B_{\circ}$}}
\def\BZTH{\hbox{$B_{{\circ},{\sf Th}}$}}
\def\GamTH{\hbox{$\Gamma_{\sf Th}$}}
\def\AZ{\hbox{$A_{\circ}$}}
\def\vZ{\hbox{$v_{\circ}$}}

\def\vE{\hbox{$v_{E}$}}
\def\vK{\hbox{$v_{K}$}}
\def\vw{\hbox{$v_w$}}
\def\rhuw{\hbox{$\rho_w$}}
\def\rhubold{\hbox{$ {\bf \rho}$}}
\def\dist{\hbox{$d$}}

\def\Teff{\hbox{$T_{\rm eff}$}}
\def\Halpha{\hbox{{\rm H}$\alpha$}}
\def\vinf{\hbox{$v_\infty$}}
\def\vesc{\hbox{$v_E$}}

\def\vparlel{\hbox{$v_{\|}$}}
\def\vphi{\hbox{$v_{\phi}$}}
\def\Bphi{\hbox{$B_{\phi}$}}
\def\vphipeak{\hbox{$v_{{\phi},{\rm peak}}$}}
\def\rhuD{\hbox{$\rho_{\rm D}$}}
\def\rhuDZ{\hbox{$\rho_{\rm D,c}$}}
\def\rhuZ{\hbox{$\rho_{\rm c}$}}
\def\NZ{\hbox{$n_{\rm c}$}}
\def\NDZ{\hbox{$n_{\rm D,c}$}}

\def\rhu1{\hbox{$\rho_{\rm 1}$}}
\def\half{\hbox{\small{$\frac{1}{2}$}}}
\def\Mdot{{\hbox{$\dot M$}}}
\def\Msunyr{\hbox{$M_\odot\,$yr$^{-1}$}}
\def\kms{\hbox{km$\,$s$^{-1}$}}
\def \etal{et~al.}
\def \ie{i.e.}

\received{2002 February 19}
\begin{document}

\title{A Magnetically Torqued Disk Model for Be Stars}

\author{J. P. Cassinelli\altaffilmark{1}, J. C. Brown\altaffilmark{2},
M. Maheswaran\altaffilmark{3}, N. A. Miller\altaffilmark{1}, and 
D. C. Telfer\altaffilmark{2}}

\altaffiltext{1}{Dept. of Astronomy, University of Wisconsin-Madison,
475 N. Charter St. Madison WI 53706; cassinelli@astro.wisc.edu, 
nmiller@astro.wisc.edu}
\altaffiltext{2}{Dept. of Physics and Astronomy, University of Glasgow,
Glasgow G12 8QQ, UK; john@astro.gla.ac.uk, deborah@astro.gla.ac.uk }
\altaffiltext{3}{University of Wisconsin-Marathon County, 518 S. 7th Av.
Wausau, WI 54401, mmaheswa@uwc.edu}

\begin{abstract}
Despite extensive study, the mechanisms by which Be star disks acquire
high densities and angular momentum while displaying variability on many
time scales are still far from clear. In this paper, we discuss how
magnetic torquing may help explain disk formation with the observed
quasi-Keplerian (as opposed to expanding) velocity structure and
their variability.  We focus on the effects of the rapid rotation
of Be stars, considering the regime where centrifugal forces provide
the dominant radial support of the disk material.

Using a kinematic description of the angular velocity, \vphi$(r)$, in the
disk and a parametric model of an aligned field with a strength $B(r)$ we
develop analytic expressions for the disk properties that allow us to
estimate the stellar surface field strength necessary to create such a disk
for a range of stars on the main-sequence.  The fields required to form a
disk are compared with the bounds previously derived from photospheric
limiting conditions. The model explains why disks are most common for
main-sequence stars at about spectral class B2 V. The earlier type stars
with very fast and high density winds would require unacceptably strong
surface fields ($> 10^3$ Gauss) to form torqued disks, while the late B
stars (with their low mass loss rates) tend to form disks that produce only
small fluxes in the dominant Be diagnostics. For stars at B2 V the average
surface field required is about 300 Gauss. The predicted disks provide an
intrinsic polarization and a flux at \Halpha\ comparable to observations.
The radial extent of our dense quasi-Keplerian disks is compatible with
typical estimates. We also discuss whether the effect on field containment
of the time dependent accumulation of matter in the flux tubes/disk can help
explain some of the observed variability of Be star disks.
\end{abstract}


\section{Introduction}\label{intro}

The presence of disks around rotating stars has long been realized from
observations, but has never been explained in a satisfactory way.
Particularly pronounced are the disks around the emission line Be stars.
These can be detected because of the strong double-peaked emission lines of
\Halpha, large IR continuum excesses, and intrinsic polarizations. Because of
these interesting and easily detectable properties, Be stars and their disks
have been intensively studied for many years (see reviews in
Smith, Henrichs, \& Fabregat (2000) and Yudin (2001)).
Nevertheless, the origins of their high densities, angular
momentum, spatial structure, and variability remain a major puzzle. In
addition to the dense equatorial disks, it is well known that Be stars have
substantial winds, roughly isotropic, with typical $\Mdot \simeq
10^{-(10 \to 8)} \Msunyr$ and terminal velocities near
$\vinf\simeq 1000~\kms$ ( Marlborough \& Peters 1986;
Grady, Bjorkman, \& Snow, 1987; Prinja 1989)

Hot stars exhibit rotational line broadening indicative of surface
equatorial rotation speeds which are high, but less than about 80 percent
the centrifugal (Keplerian) limit. Thus rotation alone is not sufficient to
centrifuge material off the equator to form a disk. Owing to the high
luminosity of the stars, wind material leaves the general stellar surface
due to radiative forces on the line opacity. Line-driving, however, only
works well where the wind material is not too optically thick, and has a
sufficient velocity gradient for the absorbing atomic lines to sweep through
the stellar continuum spectrum. Observational evidence suggests that Be star
equatorial disks are too dense and too slowly expanding for line driving
alone to support them against gravity.

One possible mechanism proposed to create disks is the Wind Compressed Disk
(WCD) model of Bjorkman \& Cassinelli (1993). In this, stellar wind matter
accelerating out from an intermediate latitude on a rotating star has a
trajectory orbital plane  which crosses the equatorial plane. The
continuous outflows of such matter from the upper and lower hemispheres
`collide' in the equatorial plane and form a compressed disk with
density enhancement by a factor of $\sim 100 $, which is on the low side for
what is needed observationally. Irrespective of other theoretical issues
raised concerning the WCD model (Owocki, Cranmer, \& Gayley 1996), the fact
is that all the observational indications seem to point toward a disk which
is mainly moving azimuthally, supported against gravity by centrifugal
forces, i.e.,~a near Keplerian disk $(v_r \ll \vphi \approx
\sqrt{GM/r})$ rather than one which is mainly outflowing $(v_r \gg
\vphi)$.

The near Keplerian and high density disk inferred from observations
poses questions which we address in this paper: How does the material
acquire more angular momentum than it possessed when it left the
stellar surface? How does such an equatorial disk accumulate material,
given that its density and small radial velocity gradient prohibit
radiative driving from overcoming gravity in the equatorial plane?
Proposed answers to these two questions have included schemes in which
matter is ejected from the star selectively in the pro-grade direction
to boost its rotation up to orbital values. In particular, Hanuschik (1999) 
considered how an outflow of particles with an isotropic
Maxwellian distribution could create an orbital regime comprising the
faster pro-grade particles with the rest falling back.  Owocki \&
Cranmer (2001) proposed non-radial pulsations with an anisotropic
radiation force favoring pro-grade gas flow. While these ideas
are interesting, the direct centrifuging action that a strong enough
stellar magnetic field could provide seems to be a more direct way of
producing the disk.

The possible role of magnetic fields in Be disk production has been
mentioned quite often (Friend \& MacGregor 1984; Poe \& Friend 1986;
Ignace, Cassinelli, \& Bjorkman 1996). These dealt with
applications of magnetic rotator wind theory, which explored the 
various degrees of control of the outflowing gas by the magnetic fields
(Belcher \& MacGregor 1976; Hartmann \& MacGregor 1982;
Poe, Friend \& Cassinelli, 1989; Mestel 1990). The magnetic rotator
models considered the effects of the field on the azimuthal motion of 
the gas, the terminal speed of the wind, and the equatorial mass loss rate.

Several papers have discussed aspects of the time-dependent
accumulation of stellar wind matter channeled by strong magnetic
fields. This has been mainly in the context of early type Chemically
Peculiar (CP) stars with very strong magnetic fields (Havnes \&
Goertz 1984; Babel \& Montmerle 1997a), but with ideas being extended
recently toward applications to other hot stars and Be star
phenomena (Babel \& Montmerle 1997b; Donati \etal\
2001). Specifically, Havnes and Goertz (1984) discussed the effect of
the growing `magneto-spheric' density and  mechanisms which might
disrupt the field, releasing energy and producing X-rays. Babel and
Montmerle (1997a) applied these ideas to develop a magnetically
confined wind-shocked model for X-rays from the Ap star IQ Aur and
extended this (Babel and Montmerle, 1997b) to the young star
$\theta^1$ Ori C. Donati \etal\ (2001) applied the model to $\beta$
Cep which has a magnetic field much lower than those on CP stars, and
which, they concluded, is highly oblique to the axis of slow rotation
of this star. All of these papers, however, neglected the contribution
of rotation to the dynamics of the magnetically channeled wind.  This
approximation cannot be valid for stars of near critical rotation
since the rotational energy density of the channeled wind, and even
more so of the shock compressed disk, may far exceed the outflow
energy density. Nor can it say anything about how disk material
acquires the quasi-Keplerian angular momentum values observed. In this
paper we therefore review the problems and address the other limiting
regime where the rotational term is dominant (centrifugal force
balancing gravity). We do so to evaluate what field is necessary to
provide sufficient torquing to produce a dense disk with
quasi-Keplerian speeds.

We make use of the fact that near main-sequence
B stars have a mass outflow and the channeling of that flow can
lead to shock compressed disk material. The relatively high densities
in that disk lead to the strong \Halpha, excess free-free fluxes and
electron scattering polarization for which Be stars are known.

We develop equations to find the field needed both to
transfer angular momentum to material that is driven up from the star by
line driven wind forces, and to redirect that flow toward the equatorial
plane. This regime of a magnetically directed flow is non-spherical and,
because there is a build up of matter, is also non-steady. The directed flow
results in the formation of a disk-like structure the azimuthal
motion of which is `quasi-Keplerian'. By this we mean that the velocity
distribution is not strictly Keplerian but the motion is primarily azimuthal
and with speeds of order the Keplerian value.

\begin{figure}
\epsscale{1.0}
\plotone{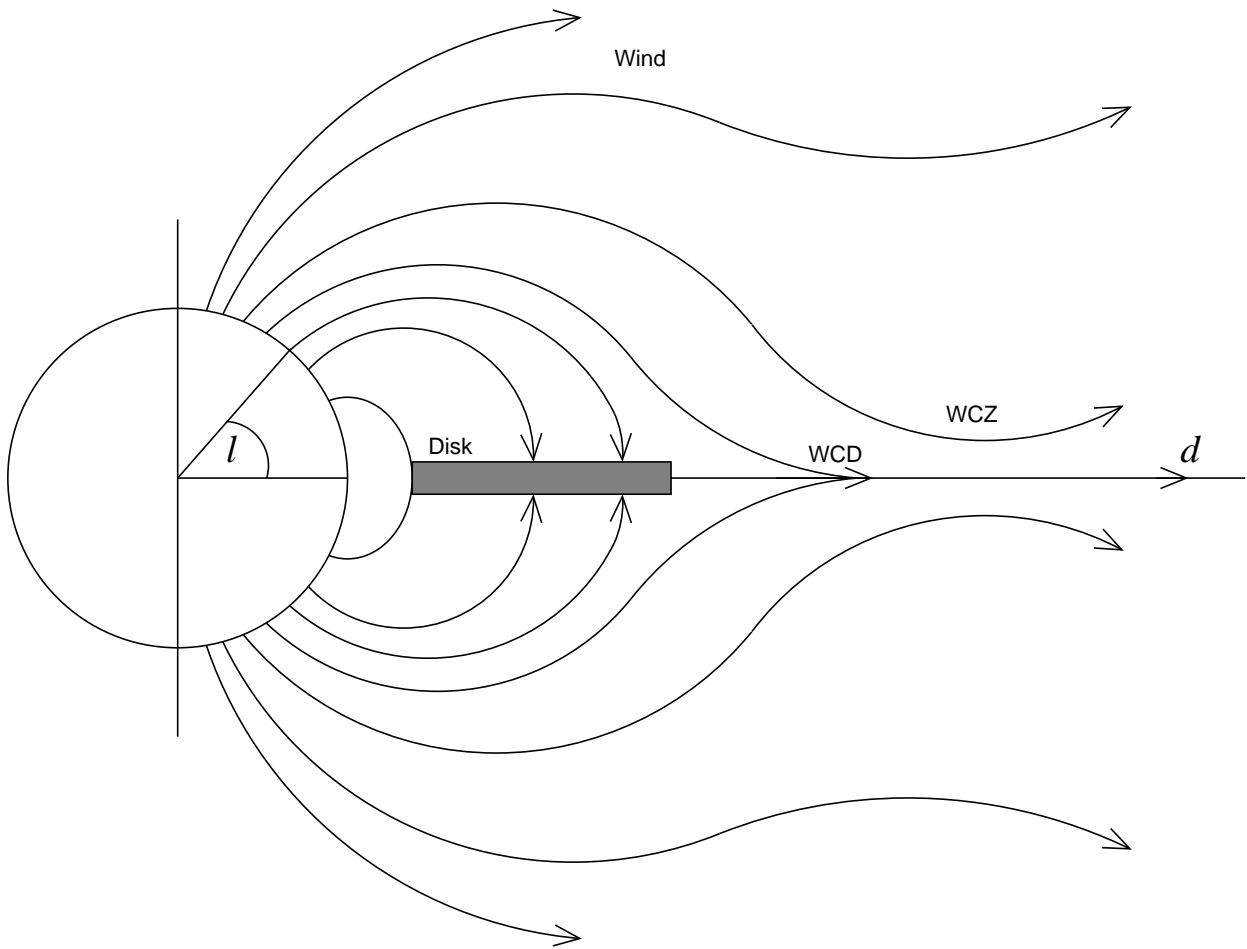}
\caption[]{Illustrates the overall structure assumed in our
Magnetically Torqued Disk model.  The star has a dipole-like inner
field with axis aligned with the rotation axis of the star. Material
leaving the star at latitude \protect\lami\ is channeled by the field
to the equatorial disk at $r=d(\protect\lami)$. The rotation causes the
channeling process to transfer angular momentum from the star to the
disk.  This will lead to a distribution of angular speed \protect\vphi\ versus
the equatorial distance \protect\dist\ from the star. Note that in the outer
regions the field lines are stretched and drawn out by the flow from
the star.}
\label{Fig:structure}
\end{figure}

We have chosen to use an aligned field of dipole-like shape for the
basic magnetic topology, $\Bvec(r,\theta)$, with the $r$-dependence
parameterized.  As for the gas, it will for the most part be assumed
to be confined by the field. Our main interest is not in the flow
trajectories $\Vvec(r,\theta,\phi)$, but rather in the azimuthal
velocity, \vphi(\dist), in the equatorial plane at radial distance,
\dist. The closed field will channel the gas to a specific radial
distance, \dist$(\lami)$, that depends on the latitude \lami\ the
field tube had at the stellar surface, the larger \lami\ corresponding
to larger \dist.  Field lines from high latitudes, $\lami \gap \pi/4$,
near the pole, would intercept the equator at very large distances,
but before reaching those distances the wind energy density will
exceed that of the field and no longer be magnetically dominated,
since even a dipole field declines as $d^{-3}$. Thus, there will be a
broad, open, polar wind that fills the envelope volume far beyond the
disk and above the closed lines of force that are channeling lower
latitude material toward the disk.  For our study, we consider both
the strictly closed field regions and the transition to the open ones
as illustrated in Figure~\ref{Fig:structure}.

The simplest picture of the situation would be an abrupt ``switch'' model,
in which the field is assumed to be fully capable of causing solid body
rotation, $\vphi(d) \propto d$, out to an \Alfven\ distance, at which
the field energy density has decreased to that in the gas flow. Beyond
that transition radius the gas would be dominant and in particular its
angular velocity would satisfy conservation of angular momentum, $\vphi
\propto 1/d$. In this representation, there is a discontinuous shift in
\vphi(\dist) (from $\vphi \propto d$ to $ \vphi \propto d^{-1}$). Although
this picture may be useful for understanding some basic concepts involved in
the torquing of disks, we know from magnetic rotator stellar wind theory
that no such abrupt shift occurs at the \Alfven\ point in either \vphi\ or
the angular momentum.

To allow for a more realistic behavior of the angular speed and
angular momentum transfer, we introduce a ``smooth transition'' model,
in which a continuous parametric function, $\vphi(d)$ is specified for
the angular velocity as a function of distance in the equatorial
plane. This function reaches a maximum, but the angular momentum
continues to increase with increasing radius by a factor of about two,
transmitted by the bent field lines. We are interested in cases in
which $\vphi(d)$ reaches at least to the Keplerian speed $\vK =
\sqrt{G M/d}$. The field lines will not be considered drawn open until
at least a distance, \dmax, where the Keplerian ratio, i.e. $\vphi(d)
/\vK(d)$ is at its maximal value. The maximal Keplerian ratio radius
is identified with the \Alfven\ radius and, from that radius outward,
we assume that the gas motion then tends to the field-free flow limit.
Thus, at large $d$, it obeys conservation of angular momentum, and the
particles remain in their orbital planes defined at the point where
they became free from the magnetic field forces.

The subsequent flow trajectories in the far field region become similar to
those in wind compression theory, either in the form of the WCD model or the
Wind Compressed Zone (WCZ) model of Ignace, Cassinelli, \& Bjorkman (1996).
The WCZ model was developed to describe the structure of a wind which has a
rotation that is too slow to form a shock compressed disk, but nonetheless
can produce density enhancements in the low latitude regions. The
enhancements in density (by a factor of 10 or so) have an effect on the
observational properties of a wide range of stars. The field lines are drawn
out in the far field regime as described by Ignace, Cassinelli, \& Bjorkman
(1998).  Neither the WCD nor the WCZ models can explain the quasi-Keplerian disks of Be stars.

The relation to WCD theory of the current model for a shock compressed
disk is as follows. The WCD model is ``kinematic'' in that the {\it
radial velocity structure} of the outflow is chosen to obey the
well-known beta law distribution, instead of being computed from the
force of radiation on line opacities. In WCD theory, the velocity
component \vphi\ is assumed to be determined by conservation of
angular momentum, and thus \vphi\ decreases rapidly with radius and
does not attain the quasi-Keplerian angular speeds inferred from
observations. In the current paper, we again use a kinematic approach,
but now specify an {\it azimuthal velocity structure}, $\vphi(d)$,
that assures quasi-Keplerian velocities are reached for a sufficiently
strong magnetic field. This kinematic approach allows us to derive
constraints on the surface field, \BZ, and the surface spin rate
parameter, \SZ, that are needed for a disk to form. Mass and magnetic
flux conservation requirements provide information regarding the disk
density distribution, the radial extent, and the timescale for
variability. Where needed we also use the wind beta velocity law to
provide an estimate of the flow velocities driven by the star's
radiation field. The derived structure for stars with ``torqued
disks'' allows us to estimate observational properties such as
\Halpha\ line strengths, intrinsic polarizations, and continuum fluxes
at a variety of wavelengths useful for comparisons of the model with
observations.

A major goal is to derive lower limits on the field strengths required for the
existence of centrifugally-supported disks produced by an aligned
dipole-like field on a Be star. As a first rough estimate, showing that the
fields required to channel a flow are not unreasonable, we consider magnetic
rotator wind theory. In that theory, as explained in Ch 9 in Lamers \&
Cassinelli (1999) (hereafter L\&C), there is a ``primary mass loss
mechanism'' (in our case the line driven wind theory), which sets the
stellar mass loss rate, and the terminal velocity.  An amplification of both
\Mdot\ and \vinf\ can be produced by the co-rotating magnetic field. A
transition occurs from the case in which the field acceleration effect is
unimportant (a ``slow magnetic rotator'') to the case in which the field
produces a strong effect (a ``fast magnetic rotator'', or FMR). This
transition occurs when the Michel velocity, \VM, (which is very close to the
terminal velocity in FMR theory) is as large as the terminal velocity owing
to the primary mass loss mechanism. The Michel velocity is related to the
Poynting flux which makes the enhanced outflow energetically possible. As a
starting point, we consider it reasonable to assume that the field necessary
to dominate the flow and channel it to the equatorial disk should be at
least comparable to the transition field strength from slow to fast magnetic
rotator theory for a particular star. The Michel Velocity, \VM, is given by
quantities defined at the surface of the star; the surface field, (\BZ),
rotation rate (\OmegZ), and mass loss rate, \Mdot. Solving for \BZ\ in the
equation for \VM\ gives a minimal field based on FMR considerations.

\begin{eqnarray}
B_{\rm FMR} &=& \frac{(\Mdot \VM^3)^{1/2}}{R^2 \OmegZ}\\
            &\approx& \frac{(\Mdot \vinf^3)^{1/2}}{\SZ \sqrt{RGM}}\\
            & =     & 6.9\units{\rm Gauss} \frac{\Mdot_{-9}^{1/2}
                     v_8^{3/2}}{\SZ (R_{12} M_{10})^{1/2}} \label{Bfmr}
\end{eqnarray}
where $R$ is the stellar radius.  We set $R \OmegZ$ equal to the
surface equatorial speed $\vZ$, which is expressed as the equatorial
Keplerian speed, \vK(R) times a fraction \SZ. We also set $\VM = \vinf
\approx \vesc$ (where \vesc\ is the escape speed at the stellar
surface). For numerical values, we express the velocity by the ratio
$v_{8} = \vinf/(10^8~ {\rm cm~ s^{-1}})$; the stellar mass and radius
ratios by $M_{10}= M/(10 \Msun)$, $R_{12} = R/(10^{12}\units{cm})$;
and the mass loss rate by $\Mdot_{-9} = \Mdot/(10^{-9} \Msunyr)$.

A roughly similar value for the magnetic field,
as in Equation~(\ref{Bfmr}), follows from requiring that the energy density
of the surface field $\BZ^2/8 \pi$ should exceed that of the gas
$\rho_{\circ} v_{\circ}^2/2$. The derived field of order
10 Gauss (for \SZ$\simeq 1$) is a very modest one. However, as we
shall see, it is a lower bound on \BZ\ because it effectively uses the
density in the wind, which is much below (\ie\ $\simeq 10^{-3}$--$10^{-4}
\times$) the densities that are produced in the centrifugally supported disk,
so that the field ($\propto \rho^{\half}$) torquing the disk is 30 to 
100 times larger than that in Equation (3).

\section{Schematic Description of the Model}\label{themodel}

Let us start by considering the case of a non-rotating star with a strong
surface magnetic field, which we take to be of dipole-like form (but of
more a general $r$-dependence), and with a
uniform wind mass flux, $ \FMZ =\dot{M}/(4 \pi R^2)$, from its surface.
If we consider starting from a vacuum environment and initiate an outflow,
gas will be channeled along the field lines
(assuming $\frac{1}{2} \rho\ v_w^2 \ll B^2 / 8
\pi$), and along with matter from the opposite hemisphere,
tend to fill the flux tubes to ever greater distances with wind compressed
gas, creating an extended dense disk. This is the non-rotating case 
considered by Babel \& Montmerle (1997a). The flux tubes would fill
toward pressure balance.  However, the resulting decrease in the velocity
gradient would probably quench radiative driving, tending to cause the
matter to fall back, although Babel \& Montmerle (1997a) suggest otherwise.
We suspect that the Rayleigh Taylor instability might be important and
prevent a wind flow from supporting a relatively dense envelope against
gravity.

When we introduce rotation, the magnetic field exerts an azimuthal
force on the gas as well as acting to constrain its radial flow path to be
along the flow tubes. As is illustrated in Figure~\ref{Fig:structure}, the
latitude
\lami\ at which the flow originates will determine the azimuthal velocity of
matter as it is channeled into the equatorial disk. Matter
originating in some latitude range $\lami_1 < \lami\ < \lami_2$ will tend to
reach the equator at $\dmin < \dist < \dmax $ with moderate speeds \vparlel\
parallel to the field lines in the corotating frame. As a result of the
influx of matter, there will be a build up of mass in these sectors of the
disk. Although compressed to high densities by the shock fronts of the
incident flows, the material in the disk will be supported against infall
mostly by centrifugal forces.

We discuss below the parameters of the various resulting regions in
the equatorial plane, and also the timescale on which they change. We
find the latter to be long compared to typical rotation periods.  So,
for the purposes of this paper, we assume we have reached a nearly
steady stage of moderately dense disk creation and consider the effect
of stellar rotation.

Most of our interest is in the collision of winds in the equatorial plane,
which involves material that is near and above the local Keplerian speed.
Therefore, we find it convenient to
express equations in terms of the dimensionless equatorial distance variable
$x=d/R$, and the variable $\SK(x)=\vphi(x)/\vK(x)$, which is the ratio
of the azimuthal speed to the local Keplerian speed $\vK(x)$.
A fundamental parameter for a specific star is the ``stellar spin rate
parameter''
\SZ, which is the ratio  $v_{\circ}/ \vK(R)$ of the stellar
rotation speed to the Keplerian velocity at the stellar equator; thus
\SZ = \SK (at $x=1$).  In principle, \SZ\ can have values in the range
$ 0 \leq \SZ \leq 1$. However, for most Be stars the range is
$ 0.6 \la \SZ \la 0.8$. We assume for simplicity that the star
is spherical and not distorted by the rotation.

\section{Existence and Extent of Magnetically Centrifuged Disk}\label{thedisk}

We begin by finding a form for $\vphi(d)$ that is consistent with known
magnetically centrifuged wind systems and provides a kinematic distribution
that we can use at all distances. Although the topology of the field in the
rotating dipole case is very different, there are properties of the
azimuthal velocity that should carry over from magnetic rotator theory
(L\&C Chapter 9). At small radii, the azimuthal speed behaves
as in the solid body case, i.e. $\vphi \propto d$, while farther out it
satisfies the conservation of angular momentum relation $\vphi\
\propto {1}/{d}$.  Now let us find a kinematic expression for
\vphi, that is valid both near and far from the star and also provides
a continuous form that asymptotically matches both behaviors.

\subsection{The Description of the Angular Velocity, 
{\boldmath $v_\phi(x)$}, in the Disk}

At small distances from the star where the field is strong, we expect
the azimuthal speed of disk matter to be controlled by centrifuging,
i.e., to follow the co-rotation distribution
\begin{equation}\label{vphidEQR}
\vphi \simeq \vZ~x \qquad (x \gtrsim 1)
\end{equation}

\noindent Farther out, the angular velocity in the region well beyond the
\Alfven\ distance should follow the conservation of angular momentum expression
$\vphi = (\vphipeak\ \dpeak)/d$, where \vphipeak\ is the maximum $\vphi(d)$,
which occurs at $\dist=\dpeak$. We express \dpeak\ in terms of the
dimensionless parameter, $\alpha$, where

\begin{equation}\label{alphdef}
\alpha  = \dpeak/R = \xpeak
\end{equation}
and thus $\vphipeak \approx \vZ \alpha$. At large $x$ we will have
\begin{equation}\label{vphidrA}
\vphi(x) \simeq \frac{\vZ\ \alp2}{x}  \qquad (x > 1, x \gg \xpeak )
\end{equation}

At intermediate equatorial distances, the exact form of $\vphi(x)$ depends
on details of how the field bends in reaction to the inertia of the matter.
However, a reasonable form for $\vphi(x)$ parameterized by $\alpha$, and
which has the limiting behaviors (\ref{vphidEQR}) and (\ref{vphidrA}), is

\begin{equation}
\label{vphi/v0}
\frac{\vphi(x)}{\vZ} = \frac{\alp2 +1}{\frac{\alp2}{\displaystyle x} +x }
\end{equation}
where the numerator normalizes $\vphi(x)$ so that $\vphi(x=1) = \vZ$. This
analytic form proves convenient for exploring the properties of
the disk driving regime. Although Equation~(\ref{vphi/v0}) is purely
empirical, it is designed to reflect the actual behavior of matter torqued
in a magnetic field.

\begin{figure}
\epsscale{0.7}
\vspace{1cm}
\plotone{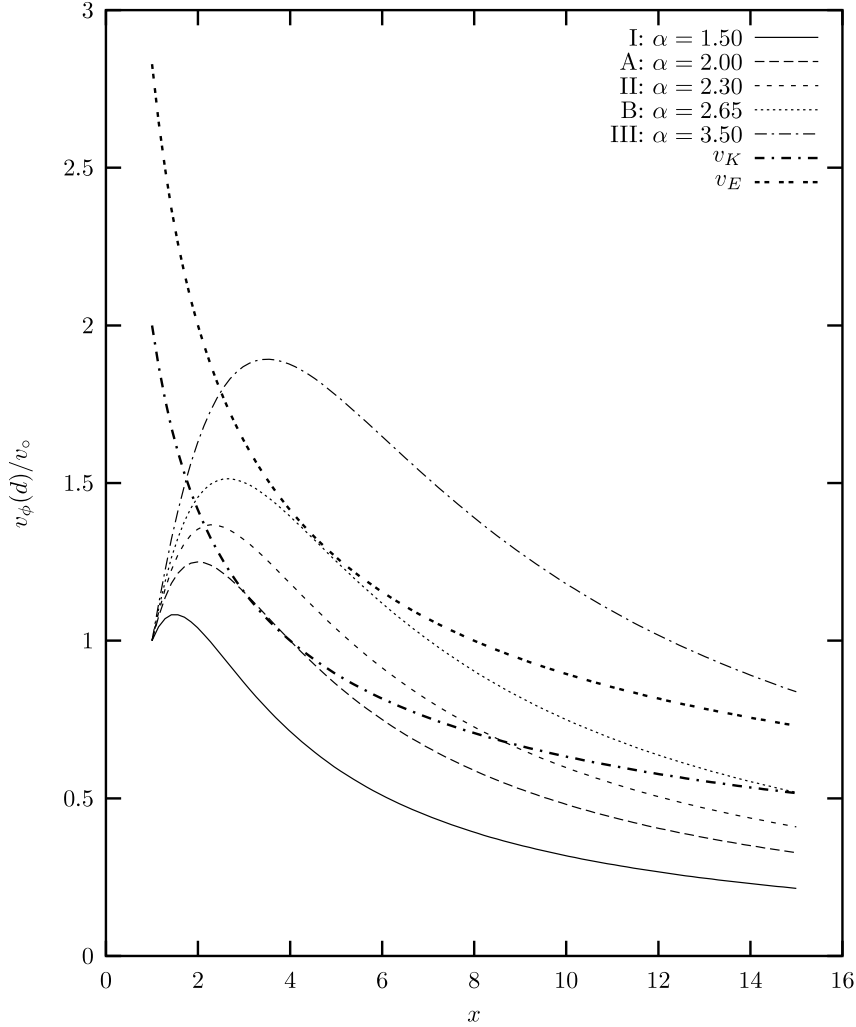}
\caption[]{The assumed parametric distribution of \protect\vphi (\protect\dist) versus
equatorial distance. These are not flow lines but rather give the
azimuthal velocity \protect\vphi (\protect\dist) of material as it reaches the
equatorial plane having started at latitude sector $\protect\lami$. A spin rate of $\protect\SZ = 0.5$, is used for this figure. In the
near solid body region near the star the \protect\vphi\ for increasing \protect\lami\
increases with distance, $d(l)$ from the star.  Note that \protect\alfa\
depends on the magnetic field, the spin \protect\SZ\ and outflow density in
the wind. Curves for five stars of different \protect\alfa\ values are shown
(from Equation~\protect\ref{vphi/v0}).  Also shown are the Keplerian $\protect\vK(d)$
distribution and the parallel but larger escape speed distribution
$\protect\vE(d)$.}
\label{Fig:vphivsx}
\end{figure}

Figure~\ref{Fig:vphivsx} shows $\vphi(x)$ from Equation~(\ref{vphi/v0}) for
various $\alpha$ values, as functions of $x$. Speeds are shown relative to
$\vZ$, so the relative height of $\vphi(x)$ and $\vK(x)$ depend on \SZ.

While it is interesting to see the dependence of the adopted
$\vphi(x)$ distribution on $\alpha$, we will see later that $\alpha$
is not a fundamental stellar parameter comparable to the rotation
quantity \SZ. In fact, for a star with a given \Mdot\ and \vinf, the
location $x=\alpha$ of the peak of the curve itself depends on \SZ\ as
well as on the field \BZ\ at the surface of the star. Only after we
explore the consequences of assuming Equation~(\ref{vphi/v0}) for
$\vphi(x)$, and particular forms for the disk field and density
distributions, will we be able to identify the various independent
stellar parameters for the Be star disk problem.

Two other curves of interest are also shown in Figure~\ref{Fig:vphivsx}:
the Keplerian
circular velocity, $\vK(x)$ and the escape speed $\vesc(x)$, where

\begin{equation}\label{vK}
\vK(x) = \frac{1}{\sqrt{2}} \vesc(x)
= \sqrt{\frac{GM}{R x}} = \frac{\vZ}{\SZ x^{1/2}}
\end{equation}

\begin{figure}
\epsscale{0.7}
\vspace{1cm}
\plotone{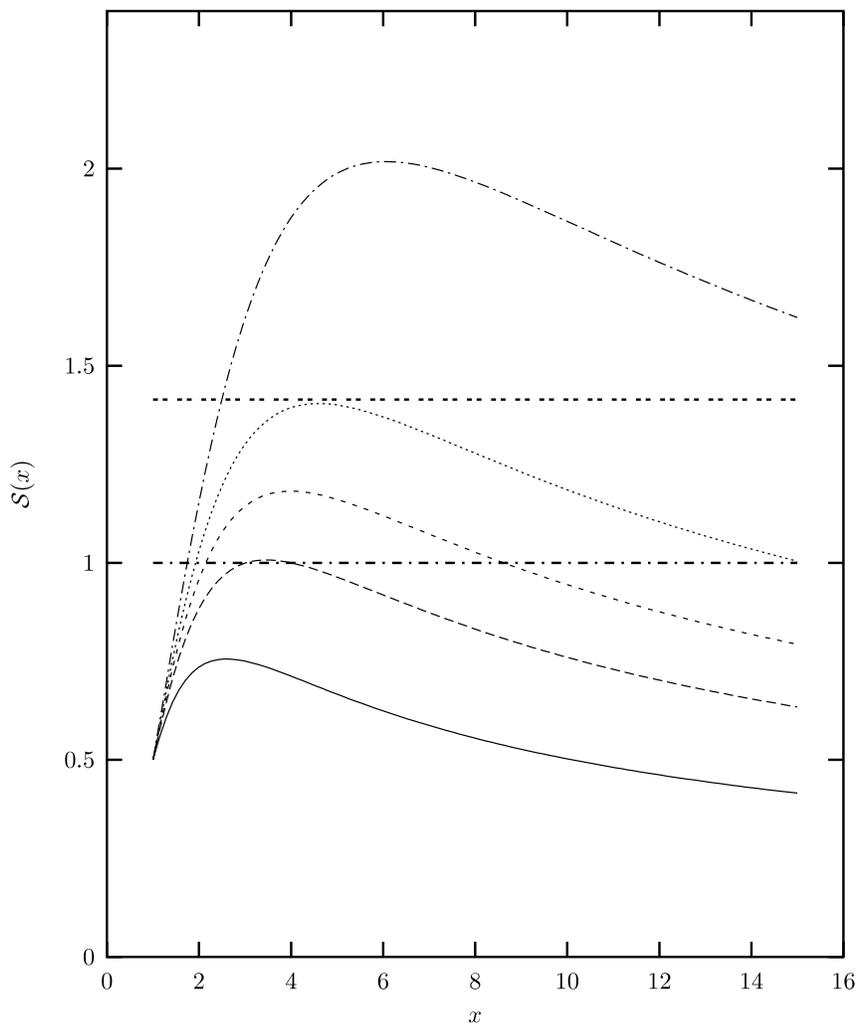}
\caption[]{The curves of $\protect\SK(x)=\protect\vphi(x)/\protect\vK(x)$ for a
range of \protect\alfa\ and for $\protect\SZ=0.5$. To contribute to a quasi-Keplerian disk
the \protect\SK\ distribution must reach values of unity and above. There is
a minimal value of \protect\alfa\ for this to occur. For higher values of
\protect\alfa, \protect\SK\ reaches a maximum farther out and that maximal point is taken
to be the outer boundary of the magnetically dominated disk.
The line types are identified in Figure 2.}
\label{Fig:Svsx}
\end{figure}

We need to compare the angular speed, \vphi\ with the local Keplerian
speed, so it is convenient to use the Keplerian ratio
($\SK(x)=\vphi(x)/\vK(x)$) introduced previously.  Using
Equation~(\ref{vphi/v0}) we get

\begin{equation}\label{SKSZ}
\frac{\SK(x)}{\SZ}=  \frac{\vphi(x)x^{1/2}}{\vZ}=\frac{(\alp2+1)
x^{3/2}}{\alp2+x^2}
\end{equation}
This reaches a maximum at $x= \sqrt{3} \alpha$.
Figure~\ref{Fig:Svsx} shows $\SK(x)$ versus $x$, for various \alfa\ values.
Of particular
interest is the value of \alfa\ for which \SK\ just tangentially reaches unity,
\ie\ for which there is just a single point at which the angular speed
reaches the Keplerian speed. This ``Threshold value'' of \alfa\ is called 
``\alpTH'', and it occurs at $\xTH =\sqrt{3} \alpTH$.
For a given star, \alpTH\ will determine the minimal magnetic field that is 
required to produce a torqued disk, i.e. one
with material having angular velocities at least as large as Keplerian
speed.

Figures \ref{Fig:vphivsx} and \ref{Fig:Svsx} reveal a number of regimes of
behavior
according to the value of $\alpha$,

\begin{enumerate}

\item[1.] For the  `weak' field regime,
curves of type $I$, with (\alfa $<$ \alpTH)
never produce a $\vphi(x)$ value as high as $v_K$ at any $x$, so $\SK < 1 $
at all $x$. The centrifugal force supplied by magnetic torquing to material
originating at any latitude  is not able to overcome gravity. We assume that
the matter will tend to fall back to the
star unless supported by other forces such as radiation pressure or ram
pressure of the inflowing accumulating gas. Field lines will already be bent
by the inertia of the gas at radial distances close to the stellar surface.

\item[2.] For a ``moderate'' field regime, for which $\alpTH \leqq \alfa
\lesssim 3$ (depending on \SZ) is of most interest in producing
quasi-Keplerian disks.  The
surface fields at intermediate \lami\ are strong enough to centrifuge gas to
equal (critical curve $A$) or exceed (curves of type $II$) the Keplerian
circular speed $\vK(x)$. This range in $\alpha$ will result in a
quasi-Keplerian disk over a range \xinner\ to \xouter. The distance \xinner\ is
the first point at which the material has a speed greater than the Keplerian
speed, and we choose \xouter\ to be the distance at which the Keplerian ratio
\SK\ is maximal and at which we can consider the field to lose its
dominance over the gas flow.

\item[3.] For the ``large'' magnetic field regime (curves of type
$III$ for which $\alfa >> \alpTH$), the magnetic torquing gives the
matter sufficient $\vphi$ alone to equal escape speed. However, the
field could still dominate and prevent outflow. Beyond some large
distance, outflow is assumed to occur with little further increase in
angular momentum and the angular velocity will vary as $
x^{-1}$. Arbitrarily large $\alpha$ could centrifuge and contain a
disk of arbitrary radial extent.  However, $\alpha$ depends on the
surface field for which there are bounds as described by Maheswaran
\& Cassinelli (1988, 1992), so there are in fact upper bounds to
$\alpha $ as well.

\end{enumerate}

We should also emphasize that even though the curves of \vphi\
in Figure~\ref{Fig:vphivsx}
have the appearance of trajectories, they are {\it not }  matter trajectories,
but rather  give the azimuthal
velocities of matter which has arrived at a particular distance $x$ in the
equatorial plane, having originated from  latitude $\lami$.

\subsection{Boundaries and Regions of the Disk}

Here we derive expressions for the limiting radii of particular interest to
the torqued disk problem. The innermost distance of interest, \xinner, is
that at which the torqued matter first attains the local Keplerian speed.
It occurs where $\SK = 1$. So by Equation~(\ref{SKSZ}), \xinner\ is
the solution of

\begin{equation}
\label{XinvsSZ}
\frac{(\alp2 + 1) \xinner^{3/2}}{\alp2+\xinner^2}= \frac{1}{\SZ}
\end{equation}

\begin{figure}
\epsscale{0.7}
\vspace{1cm}
\plotone{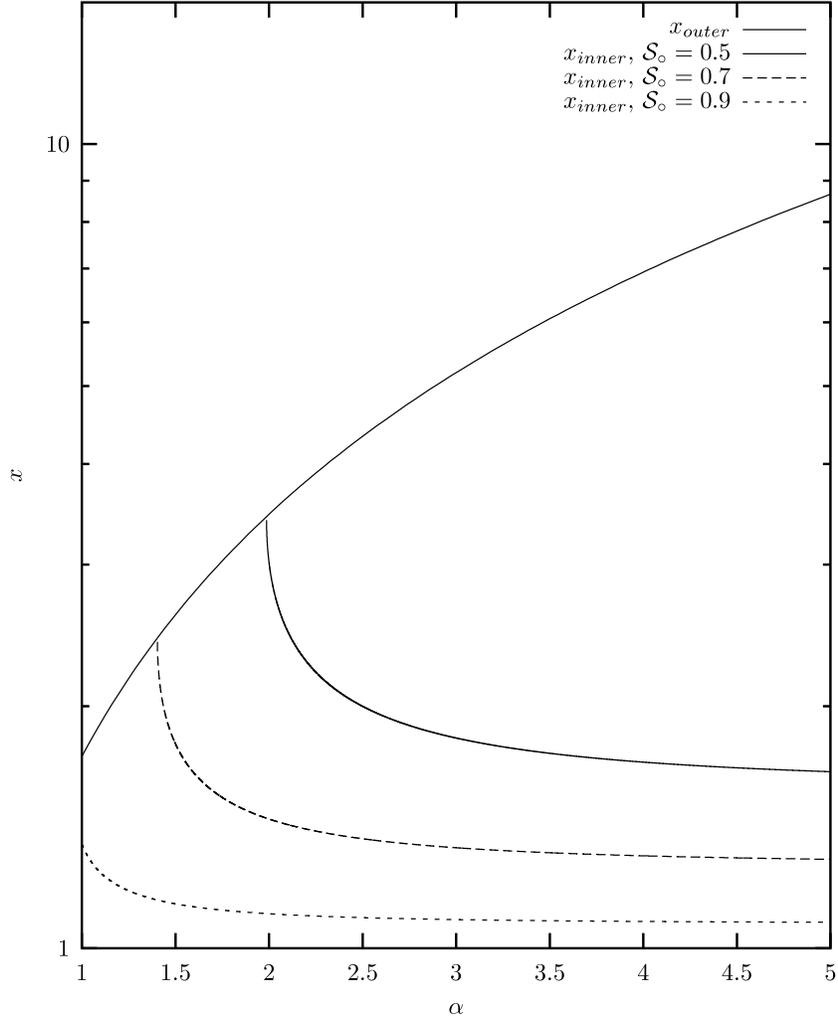}
\caption[]{The minimal and maximal radii \protect\xinner\ and \protect\xouter\ as
functions of $\protect\alfa$.  Equation~\ref{XinvsSZ} is used to calculate
\protect\xinner\ with the indicated values of \protect\SZ, while  $\protect\xouter = \sqrt{3} \protect\alfa$.}
\label{Fig:xTHvsSZ}
\end{figure}

This non-linear equation can be solved by a Newton iteration, which
converges quickly when we start with the large $\alpha $ solution to
Equation~(\ref{XinvsSZ}), \ie\ $\xinner \simeq \SZ^{-2/3}$.
Figure~\ref{Fig:xTHvsSZ} shows $\xinner(\alpha,\SZ)$ versus $\alpha$
for various \SZ, (0.5, 0.7, 0.9).

The next distance of interest is where \vphi\ peaks, namely $\dpeak =
\xpeak\ R$, which we have already seen is given by

\begin{equation}
\label{Xpeakdef}
\xpeak = \alpha
\end{equation}
and at which the rotation speed ratio is
\begin{equation}
\label{vphiSZ}
\frac{\vphi(\xpeak)}{\vZ}= \frac{\alp2+1}{2 \alpha}.
\end{equation}

To define an outer radius of the torqued
disk we need a criterion to decide where the quasi-Keplerian disk
torquing effectively ceases. For this we adopt the distance at which the
ratio $\SK(x)$ maximizes,
\ie\ where $d\SK/dx = 0$, for a fixed $\alpha$. From differentiation
of Equation~(\ref{SKSZ}) this is found to be

\begin{equation}\label{xouter}
\xouter = \sqrt{3} \alpha
\end{equation}
The value of \vphi\ at \xouter\ is given by

\begin{equation}
\frac{\vphiXout}{\vZ} = \sqrt{3}~\frac{\alp2 +1 }{ 4 \alpha}
\end{equation}
which for large $\alpha$ varies as $\sim 0.43 \alpha$. Thus, both the
location and velocity at the peak increase roughly linearly
with the parameter $\alpha$.
Note that although \xouter\ is not explicitly dependent on \SZ,
it depends on \SZ\ through $\alpha(\SZ)$.

\begin{figure}
\epsscale{1.0}
\vspace{6cm}
\plotone{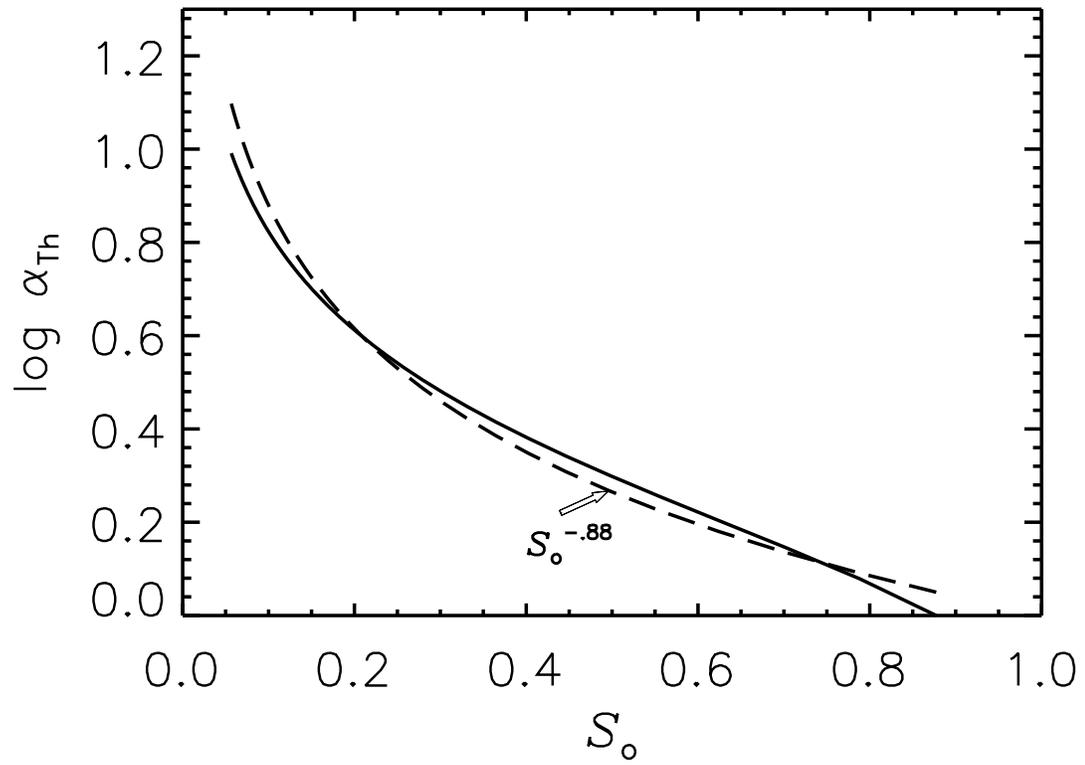}
\caption[]{The solid line shows the dependence on the spin parameter \protect\SZ\ of
the minimal value of $\protect\alpha$ (= \protect\alpTH) required for disk formation, as
given by Eq. (\ref{alphaTH}). The dashed line illustrates the good fit given
by the power law expression described by Eq. (\ref{alTHhat}). }
\label{Fig:alpTHSZ}
\end{figure}

A  key issue is the minimum value of $\alpha$ for which
a Keplerian regime exists at all.  There is a single point at
which the angular speed just reaches the Keplerian speed. Setting
$\SK = 1$ at $x =\sqrt{3} \alfa$  in Equation~(\ref{SKSZ}) 
gives

\begin{equation}
\label{alphaTH}
\alpTH^{3/2} + \frac{1}{\alpTH^{1/2}} = \frac{4}{3^{3/4}\SZ}
\end{equation}
This minimal or ``threshold'' value of \alfa, produces a Keplerian
value of \vphi\ at the singular distance $\xTH = \sqrt{3} \alpTH$.  In
Figure~\ref{Fig:alpTHSZ} we show the numerical solution for
\alpTH(\SZ) obtained from Equation~(\ref{alphaTH}).
This equation shows that \alpTH\ will be less than 1 for
values of \SZ\ larger than $2/(3^{3/4}) \approx 0.9$ (also shown in
Figure~\ref{Fig:alpTHSZ}).  However, it should be noted that at very
large spin rates the star will be substantially non-spherical and our
whole treatment becomes rather approximate. Our results should be
regarded as most reliable for values of \SZ\ that are less than 0.9.

We observe from Figure~\ref{Fig:alpTHSZ} that for the \SZ\ range most 
relevant to Be stars, an excellent fit to the solution is

\begin{equation}
\label{alTHhat}
\alpTH   \approx \SZ^{-K}; \qquad {\rm where}~~ K \simeq 0.88
\end {equation}
shown as the dashed line in Figure~\ref{Fig:alpTHSZ}.

To relate $\alpha $ to absolute scales of $B, \rho$, and \SZ,
we will  adopt $x=\xouter = \sqrt{3} \alpha$ as the point where
$B^2/8 \pi \simeq\ \rho \vphi^2/2$, \ie\ we take \xouter\
to be the \Alfven\ distance. Therefore, our definition of the \Alfven\ distance
leads to the relation

\begin{equation}\label{XAdef}
\xalfven =  \xouter 
= \sqrt{3} \alpha \qquad {\rm for}~~ \alfa \geqslant \alpTH
\end{equation}
Thus, the \Alfven\ distance is the outer boundary of our torqued disk.

We recognize that there is a degree of arbitrariness with this
definition of the \Alfven\ distance. However, just as we were guided
by wind theory in our choice of the kinematic expression for the
angular velocity, we can also get some insight from magnetic rotator
winds regarding an appropriate definition of the \Alfven\ radius. In
the case of winds the \Alfven\ radius does not correspond to the peak
of the $\vphi(x)$ distribution, but as seen in L\&C, Fig 9.3, occurs
farther out because the field continues to contribute to the speed of
the gas. The peak of our \SK\ curve (at $x= \sqrt{3} \alpha$) clearly
lies farther out than the maximum in the \vphi\ curve (at $x =
\alpha$). It corresponds to an extremum in the torquing of
the disks, in that no further increase in the Keplerian ratio
occurs. However, perhaps the best reasons for choosing the maximum in
\SK\ to be the outer boundary of a disk are seen by considering the
threshold case in which the \SK\ curve just tangentially reaches
unity, which occurs for $\alpha = \alpTH$ and at the location,
\xTH. By definition of ``a threshold distance'' we would want
the inner and outer boundaries to coincide, and the condition $\xinner =
\xouter$ occurs at \xTH. The tangentially Keplerian point is also an
important location because the components of the total velocity ($v_r$,
$v_\theta$,$ v_\phi$) are fully definable for the first time at \xTH: The
radial velocity, $v_r$ equals zero at the equatorial extremity of the
dipolar field.  The polar velocity $v_{\theta}$ is there nulled by the
collision between the oppositely directed streams. The tangential velocity
there is $\vphi =\vK(\xTH)$, so $v_{total} (\xTH) = \vphi(\xTH)$.
Furthermore, using shock conditions we will find the distance \xTH\ to be
the first radial distance in the equator for which our marginal model leads
to a quasi-Keplerian disk density (\rhuD).  Therefore, all the quantities
needed for giving an equality between the total speed and an \Alfven\ speed,
$B/\sqrt{4 \pi \rho}$, are definable at \xTH. Thus we will be able to use
the conditions at \xTH\ to define a threshold surface field \BZTH\ required
to produce a torqued disk.

Choosing the peak of \SK\ to define the outer boundary of the
torqued disk and also the \Alfven\ radius will allow us to
predict the dependence of observational fluxes on field
strengths and intrinsic stellar properties. 
We will then be able to test the resulting models against general Be
and hot-star observational requirements.

We assume that, beyond \xouter (= \xalfven), the gas flow will tend to
draw the field outward and change from being predominantly an
azimuthal flow to dominantly an outflow in the form of field-free WCD
or WCZ trajectories. The flow will make the transition to a nearly
radial wind as the latitude of the point of origin increases.  For
$\alpha > \alpTH$, the `Keplerian disk' regime extends from $ d =
\xinner\ R$ (Equation~\ref{XinvsSZ}) to $d = \xouter R$
(Equation~\ref{xouter}). Beyond the distance $\xouter= \sqrt{3} \alpha$, the
flow is no longer controlled by the field and there is a transition
from a quasi Keplerian to an outflow dominated velocity structure.

To get numerical values for these various boundaries in $x$, we need
to specify how the magnitude of $B(x)$, and $\rho(x)$ vary with
distance, for a given value of \SZ. Since $\alpha$ depends on \SZ, as
well as on $B$ and $\rho$, we specify all of these quantities in
Section \ref{Bdrhod}.  To get a rough idea at present of the expected disk
extent, we take \SZ=0.7 and assume that $B$ and $\rho$ are such that
$\alpha = 2$.  Applying Equations~(\ref{XinvsSZ}) and~(\ref{xouter}),
we find the range of the quasi-Keplerian disk to be $1.45 R
\lesssim d \lesssim 3.46 R$; a plausible range for Be star disks.

\section{The Distribution of $B(x)$ and \rhubold$(x)$ in the Equatorial Disk}
\label{Bdrhod}

\subsection{Distribution of $B(d)$}

We choose to represent the field in the system as having a topology similar
to the dipole limit, but to allow for greater generality we use a parameterized
rate of decline with distance \dist. Specifically we set

\begin{equation}\label{Bbdex}
B(d) = \BZ \left(\frac{d}{R}\right)^{-b} = \BZ~~x^{-b} \qquad (1 \leq x \leq
\xouter)
\end{equation}
where $\BZ=B(R)$ is the value at the stellar surface.  In fact,
the disk does not extend down to $x=1$ but only to $x=\xinner$.
Beyond $x=\xouter$ the field lines tend to stretch and break open and
we do not attempt to describe this here as it is not inside the torqued
disk regime.  As we will show, the field within the disk determines
its radial extent.  We also need to know the shape of the field
lines in the region out of the equatorial plane. This is because it is the
frozen-in {\it tube geometry} between the footprint of the field at
stellar latitude $l$, and the point $x(l)$ where the field crosses the
disk which determines the {\it flow geometry} (\ie\ area of the tubes) 
through which the wind mass flux is channeled.

To get the tube geometry, we make the simplifying approximation that
the field strength (and also the wind mass flux) is uniform over the
surface of the star, \ie\ \BZ\ is assumed to be the field at all $l$,
and the field lines are normal to the stellar surface. These
assumptions {\em ---} as well as constant radial \Bvec\ and constant
mass flux {\em -} will not be strictly true in a rapidly rotating
star, but they should be adequate to describe the limited range of
latitudes $l$ that adds matter to the range \xinner\ to \xouter\ in
the disk.

With these assumptions regarding the field geometry, the outgoing cross-section
of a flux tube of area \AZ\ at the stellar surface has a value $A$ at $d$ in
the equatorial plane determined by field line conservation to be

\begin{equation}\label{tubeA}
A(d)= \AZ\ \frac{\BZ}{B(d)} = \AZ x^b
\end{equation}
where we have used Equation~(\ref{Bbdex}) for the ratio of the fields at the
base and at the point of intersection with the disk.

\subsection{The Density Structure, {\bf $\rho_D$}(d), of the Disk}

We assume that the density is determined by pressure balance in the post
shock region where the flows from the upper and lower hemispheres collide.
Thus, the density near the equatorial plane is determined in much the
same way as in Bjorkman \& Cassinelli (1993), but the material is not
flowing outward, in contrast with the WCD model.  Also, since the material
has an angular velocity at or above the Keplerian speed, it is supported
against gravity by centrifugal forces. Thus the dynamic pressure does not
also need to support the gas against gravity, which is the case for the
models of the slowly rotating chemically peculiar stars
(Babel \& Montmerle 1997a).

The opposing wind streams provide a ram pressure that must be balanced
by the static gas pressure in the shock compressed region.  The
density in the shock compressed disk can be significantly larger than
that in the incident winds, especially in the near-equator zone where
the shocked gas has cooled radiatively.  We can assume the equatorial
disk region to be cool ($\approx 10^4$ K) because such a temperature
can be maintained from a radiative equilibrium balance with the
incident stellar radiation field. That is, the density is sufficiently
high that radiative cooling is effective and sufficient to balance the
heating by the stellar light.  For a Be star, the appropriate
temperature range is from 1~{\rm to} $2 \times 10^4 $\units{K},
and with the corresponding
sound speed $c_S = (kT/\mu m_H)^{1/2}$, where $\mu m_H$ is the mean
particle mass. We find by equating the disk gas pressure to the wind
ram pressure that,

\begin{equation}
\label{rampressure}
\rhuD(x) c_S^2 = \rhuw(x) \vw^2(x) =  \FM(x) \vw(x)
\end{equation}
where \rhuw\ and \vw\ are the density and velocity of the flow at the
equatorial end of the magnetic tube, and $\FM(x)$ is the mass flux there.
We use the subscript, $w$ (for wind), on both the incident velocity and
density.
We assume that the radiation forces drive a field-channeled flow
with a parallel velocity, $\vparlel$ which is
comparable to the wind speed that could have been reached at a radial
distance $d$ in the line driven wind of a non-magnetic B-star. We take
this to be given by the beta law expression

\begin{equation}\label{betalaw}
\vw(\dist)=\vinf \left(1 - \frac{R}{d} \right)^{\beta}
\end{equation}

The mass loss rate of the B-star sets the surface mass flux scale \FM,
and as stated above it is sufficient for our purposes to assume the
mass flux rate $\FMZ=\FM(x=1) $ is the same at all latitudes on the
star. The mass flux at $x$ is related to the mass flux entering at the
star via $\FM(x)\times\ A(x)=\FMZ \times \AZ$, where $A$ is the
area perpendicular to the flux tubes, as given in (\ref{tubeA}).  Thus
we obtain,

\begin{eqnarray}\label{FMx}
\FM(x) &=&\frac{\Mdot}{4 \pi R^2}~ x^{-b}\\
       &=& 5.02 \times 10^{-9} (\units{gm cm$^{-2}$ s$^{-1}$})
        \frac{\Mdot_{-9}}{R_{12}^2}~ x^{-b}
\end{eqnarray}
Using Equations~(\ref{betalaw}) and~(\ref{FMx}),
the disk density at the equator is given by the pressure balance
equation (\ref{rampressure}), thus

\begin{eqnarray}
\rhuD(x) & =      & \frac{\Mdot \vw(x)}{4 \pi R^2 c_S^2}~ x^{-b} 
                    \label{rhoin} \\
         & =      & \frac{\Mdot}{4 \pi R^2 \vinf} 
	           \left(\frac{\vinf}{c_S}\right)^2 (1-1/x)^{\beta}~ x^{-b}\\
	 & =	& \rhuZ \left(\frac{\vinf}{c_S}\right)^2
        	(1-1/x)^{\beta}~ x^{-b}\\
	 & =      & 3.04 \times 10^{-13} \units{gm}\units{cm}^{-3}
	        \frac{\Mdot_{-9} v_8}{R_{12}^2 T_4} (1-1/x)^{\beta}~x^{-b}
		\label{rhoinval}
\end{eqnarray}
where we let
\begin{equation}
\rhuZ \equiv \Mdot/(4 \pi R^2 \vinf)
= 5.02 \times 10^{-17} {\rm gm~cm}^{-3} \frac{\Mdot_{-9}}{R_{12}^2 v_8}
\label{rhuzdef}
\end{equation}
be a characteristic stellar mass loss density, and 
\begin{equation}
\rhuDZ \equiv \rhuZ (\vinf/c_S)^2 = 6.06 \times 10^3 \rhuZ 
\left(\frac{v_8^2}{T_4}\right)
\label{rhuDzdef}
\end{equation}
be the characteristic density in the disk. The corresponding
number densities are given by  $\NZ~=~\rhuZ/(\mu m_H)$ and 
$\NDZ~=~\rhuDZ /(\mu m_H)$.

From differentiation of Equation~(\ref{rhoin}), we find that
the disk density peaks at $x=1+ \beta/b \leq 1+ \beta/3$ which
lies close to  \xinner, \ie 

\begin{equation}\label{rhoDval}
\rhuD_{\sf{MAX}} = \rhuDZ\
\frac{b^b \beta^{\beta}}{(b+\beta)^{b+\beta}}
=3.04 \times 10^{-13} \frac{b^b \beta^{\beta}}{(b+\beta)^{b+\beta}}
 \frac{\Mdot_{-9} v_8}{R_{12}^2 T_4}
\end{equation}

The disk density in  Equation (\ref{rhoinval}) is nearly 4 orders
of magnitude larger than that in the ``wind'' incident on the disk

\begin{eqnarray}
\label{rhowind}
\rhuw(x)& =& \frac{\FM(x)}{\vw} \simeq \frac{\Mdot}{4 \pi R^2
\vw(x)}~x^{-b}=\rhuZ \frac{\vinf}{\vw(x)}~ x^{-b}\\
        & =& 5.02 \times 10^{-17} {\rm gm~ cm}^{-3}
	\frac{\Mdot_{-9}}{R_{12}^2 v_8} (1-1/x)^{-\beta}~ x^{-b}
\end{eqnarray}

Thus the two adjacent densities (post- and pre-shock) given by
Equations (\ref{rhoinval}) and (\ref{rhowind}) respectively, differ by
about a characteristic Mach number squared $(\vinf/c_S)^2$, as is
typically the case for isothermal shocks. In the case of early-type
stars such as the Be stars, the Mach number factor is quite large.
Finally, the large density contrast between the region in and out of
the shocked disk justifies the neglect of gas pressure compared with
the centrifugal support, as assumed in our model.

\section{The Magnetic Field for Disk Production in Terms of Stellar
Parameters}\label{diskgeneration}

\subsection{The \Alfven\ Distance and the Minimal Field to Produce a Disk}

We now have explicit parameterizations of the disk field, $B(x)$
(Equation~\ref{Bbdex}), density $\rhuD(x)$ (Equation~\ref{rhoin}), and
azimuthal velocity (Equation~\ref{vphi/v0}) as functions of $x$. Given
these, we can explicitly find the minimum or threshold surface field, 
\BZTH, needed
to torque a quasi-Keplerian disk, and the extent of the disk, as
functions of the stellar spin rate, \SZ, and wind properties, \Mdot,
$\beta$, and $v_\infty$. The existence of unique solutions for \BZTH\
and for disk inner and outer radii depends on the facts that
$\vphi(x)$ increases with $x$ in the inner region and so can exceed
\vK\ if $B$ is strong enough, and that $B^2(x)/4 \pi \rhuD$ declines
monotonically with $x$, so that the torquing effect of the field turns
off at some sufficiently large $x$. It is easy to confirm that
Equations~(\ref{Bbdex}),~(\ref{rhoin}), and~(\ref{vphi/v0}) satisfy
this monotonic decline in energy density ratio.

\begin{figure}
\epsscale{0.7}
\vspace{1cm}
\plotone{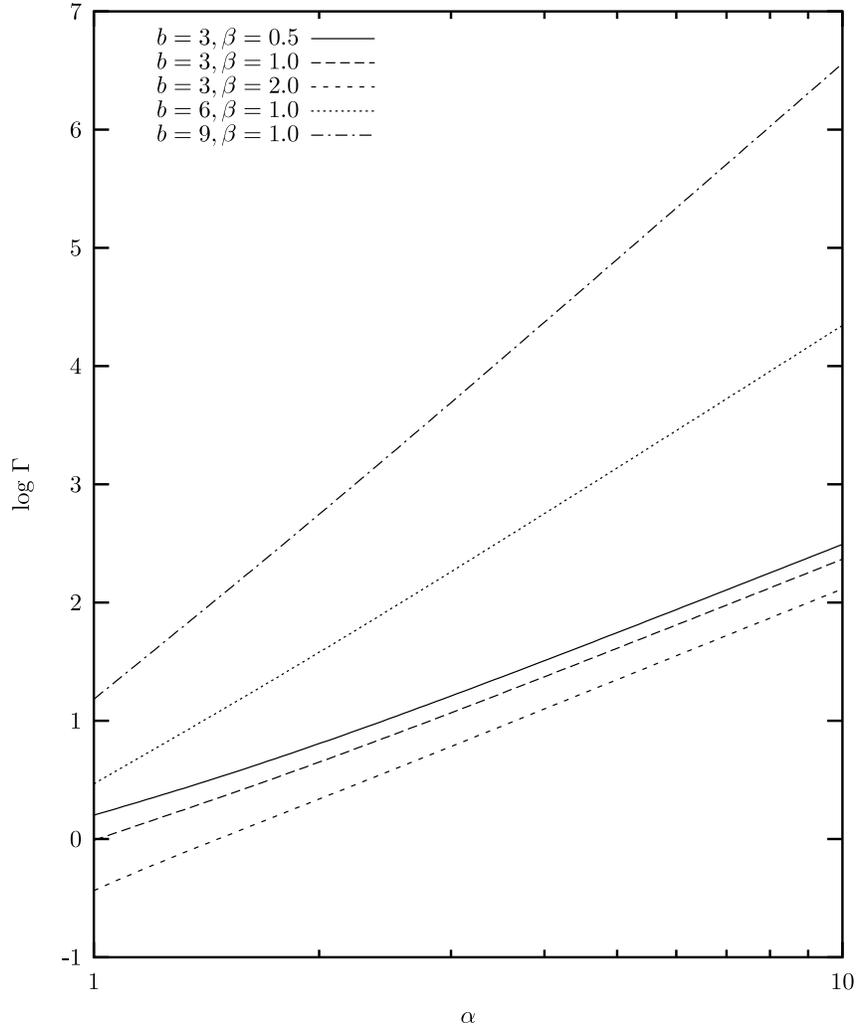}
\caption[]{The function $\protect\Gamma(\protect\alpha)$ (Eq.~(\ref{Gamdef})) is shown for the values of
$b$ and $\protect\beta$ as described by the different line styles in the legend.  The case
$b=3$ corresponds to a dipole magnetic field.}
\label{Fig:gammaalpha}
\end{figure}

To express $\alpha $ explicitly in terms of
the physical parameters of the star, we use the fact that at the rotational
\Alfven\ point, \xalfven (=\xouter),

\begin{equation}
\label{alfpoint}
\frac{B^2(\xalfven)}{8 \pi} = \half \rhuD(\xalfven) \vphi^2(\xalfven)
\end{equation}
which, on using Equations~(\ref{Bbdex},~\ref{rhoin}, and~\ref{vphi/v0}) implies
that

\begin{equation}\label{xabigequ}
\xalfven^{b+2} \left( 1-\frac{1}{\xalfven}
\right)^{\beta}\frac{(\alp2+1)^2}{(\alp2+\xalfven^2)^2} =
\frac{\BZ^2}{4 \pi \rhuZ} \frac{R}{GM}
\left(\frac{c_S}{\vinf}\right)^2 \frac{1}{\SZ^2} \\
\end{equation}
Using $\xalfven=\sqrt{3} \alpha$ (Equation~(\ref{XAdef})), and 
substituting this in Equation~(\ref{xabigequ}) gives us

\begin{equation}\label{XABdef}
\frac{3^{\frac{b}{2}+ 1}}{16} \alpha^{b-2} (\alp2+1)^2 \left( 1
-\frac{1}{\sqrt{3} \alpha} \right)^{\beta} = \Gamma^2
\end{equation}
where we use the definition

\begin{eqnarray}\label{Gamdef}
\Gamma^2 & \equiv &
\frac{\BZ^2}{4 \pi \rhuZ} \frac{R}{GM}
\frac{c_S^2}{\vinf^2}\frac{1}{\SZ^2} \\
         &=& \frac{\gamma^2}{\SZ^2}\frac{c_S^2}{\vinf^2}
\end{eqnarray}
Equation~(\ref{XABdef}) is our basic equation for relating \alfa, the
fundamental quantity in our kinematic model, to the stellar field and
basic stellar properties contained in $\Gamma$.  In
Figure~\ref{Fig:gammaalpha}, we show $\Gamma(\alpha)$ from
Equation~(\ref{XABdef}) for several $b$ values (3, 6, 9) with $\beta=1$,
 while illustrating a range of $\beta$ values (0.5, 1.0, 2.0) for the
pure dipole-field case ($b=3$).  Note that

\begin{equation}\label{gamdef}
\gamma = 
\left[\frac{\frac{\BZ^2}{8 \pi}}{\frac{GM\rhuZ}{2 R}} \right]^{\half}.
\end{equation}
Therefore, $\gamma$ is a measure of the magnetic energy density to
gravitational energy density of the wind material near the stellar
surface, and depends only on $\BZ$, $M/R \times \Mdot/\vinf$. Thus
$\gamma$ is essentially the ratio of the surface \Alfven\ speed at the
density of the wind to the surface Keplerian speed. On the other hand,
\SZ\ is simply the ratio of surface rotation speed to surface Kepler
speed, so the right-hand side of Equation~(\ref{XABdef}) (\ie\
$\Gamma$) involves only velocity ratios at the star while the
left-hand side contains only the model parameter $\alpha$ plus the
wind $\beta$ value, which is known $(\ie \simeq 1.0)$, and $b$ which
parameterizes the rate of decline of the field and we know $b \geq
3$. Note in particular that we can identify, $\gamma$,
$\vinf/c_S$ and \SZ\ as independent and fundamental
parameters, whereas the kinematic description parameter \alfa\ is not
fundamental, as  $\alpha=\alpha(\gamma,\vinf/c_S,\SZ)$.

\subsection{The Fundamental Stellar Parameters for Disk Production}

We can now translate the analysis of the disk, in \S 3,
in terms of kinematic parameter $\alpha$ into the real physical 
parameters of different stars.

First of all, we can find the surface field \BZTH\ needed to create a
quasi-Keplerian disk for a star of given $\SZ, M, R, \Mdot$, \vinf,
and $\beta$. To do so we use the results of Figure~\ref{Fig:alpTHSZ}
(from Equations~\ref{alphaTH} and~\ref{alTHhat}) to obtain \alpTH(\SZ)
for the given \SZ\ value.  We can then use \alpTH\ in
Equation~(\ref{XABdef}) (with the chosen values of the parameters 
$b$, and $\beta$) to obtain $\GamTH(\SZ)$ which is a convenient intermediate
variable encompassing all of the velocity ratios of the system.

Given $\GamTH$ from \alpTH\ for the chosen $\beta$ and $b$
and using Equation~(\ref{rhuzdef}) for \rhuZ,  we can now find

\begin{eqnarray}\label{BZTH}
\BZTH &=& \SZ~ \GamTH(\SZ)
               \left( \frac{GM \Mdot \vinf}{R^3 c_S^2} \right)^{\half}\\
       &=& \SZ~ \GamTH(\SZ) \times Y_*
\end{eqnarray}
where $Y_*$ is determined by stellar properties.

Using the reference values for $M$, $R$, \Mdot, and \vinf\
introduced for Equation~(\ref{Bfmr}) while introducing
$c_S=1.3\times 10^6 T_4^{1/2}$, Equation~(\ref{BZTH}) can be expressed
numerically as

\begin{equation}
Y_*= 71 ~ {\rm Gauss}
\left(\frac{M_{10} \Mdot_{-9} v_8}{R_{12}^3 T_4}\right)^{\half}
\label{Ystarval}
\end{equation}

\begin{figure}
\epsscale{1.0}
\vspace{6cm}
\plotone{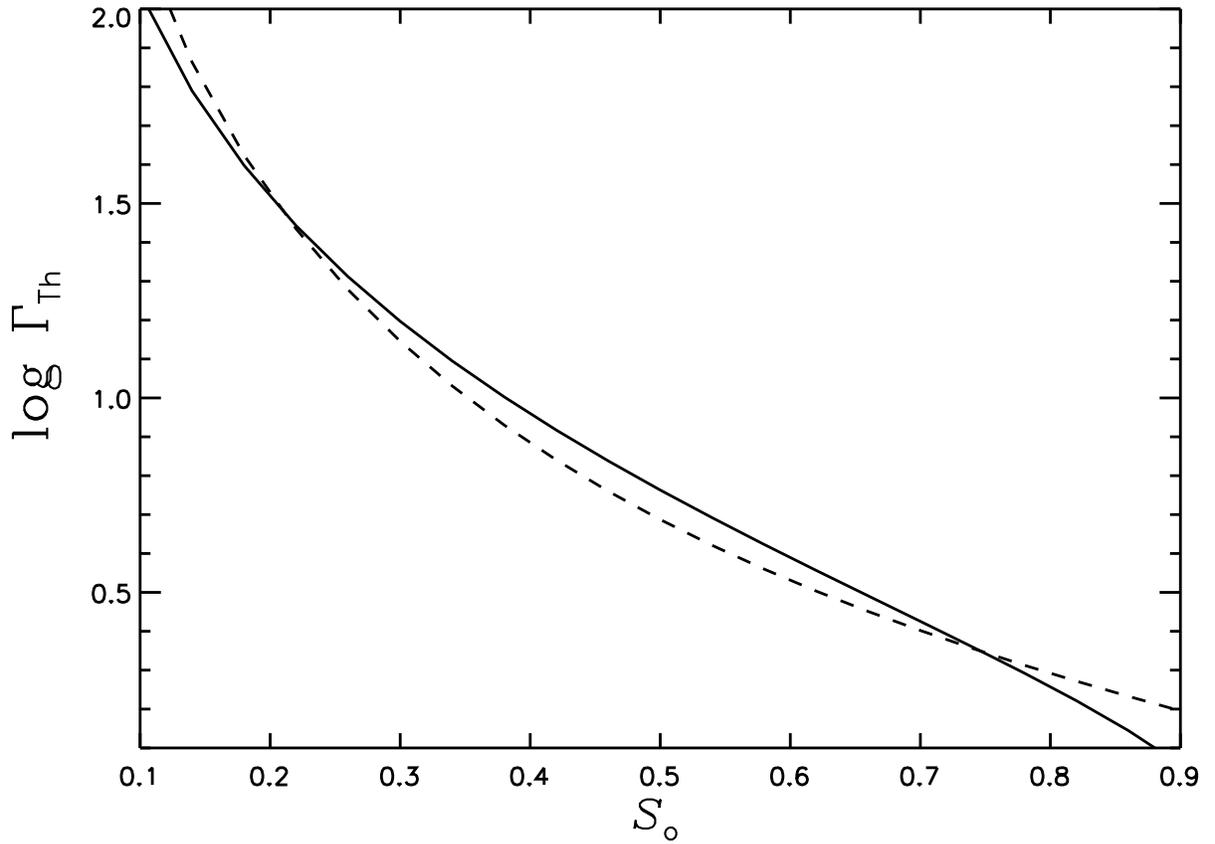}
\caption[]{The threshold value of  $\protect\Gamma$ versus \protect\SZ\
for $b=3$ and for $\beta =$ 1.0, from empirical
Equation~(\ref{Gammin1}) (dashed line) and also from the exact
solution of Equations~(\ref{alphaTH}) and~(\ref{XABdef}) (solid line).}
\label{Fig:GamTHSZ}
\end{figure}

In principle we should solve Equation~(\ref{alphaTH}) to get \alpTH(\SZ)
and hence $\GamTH(\SZ)$, but we noted that Equation~(\ref{alTHhat})
($\alpha=\SZ^{-K}$) yields an excellent empirical fit to \alpTH(\SZ).
Using the empirical fit in Equation~(\ref{Gamdef}) gives

\begin{equation}\label{Gammin1}
\GamTH(\SZ) = \frac{3^{b/4+1/2}}{4} \SZ^{-K (b/2-1)} (\SZ^{-2K}+1)
\left( 1-\frac{\SZ^K}{\sqrt{3}} \right)^{\beta/2}
\end{equation}
which we show in Figure~\ref{Fig:GamTHSZ} along with the exact solution
of Equation~(\ref{alphaTH}).

Inserting $K=0.88$  explicitly in Equations~(\ref{BZTH})
and~(\ref{Gammin1})  for the range of \SZ\ of greatest interest
for Be stars we get,

\begin{equation}
\label{BZTHval}
\BZTH(\SZ) =17.5 \units{Gauss}
\times 3^{b/4+1/2} \SZ^{1.88-0.44 b} (\SZ^{-1.76}+1)
\left(1-\frac{\SZ^{0.88}}{\sqrt{3}} \right)^{\beta/2}
\left(\frac{M_{10} \Mdot_{-9} v_8}{R_{12}^3 T_4}\right)^{\half}
\end{equation}

\begin{figure}
\plotone{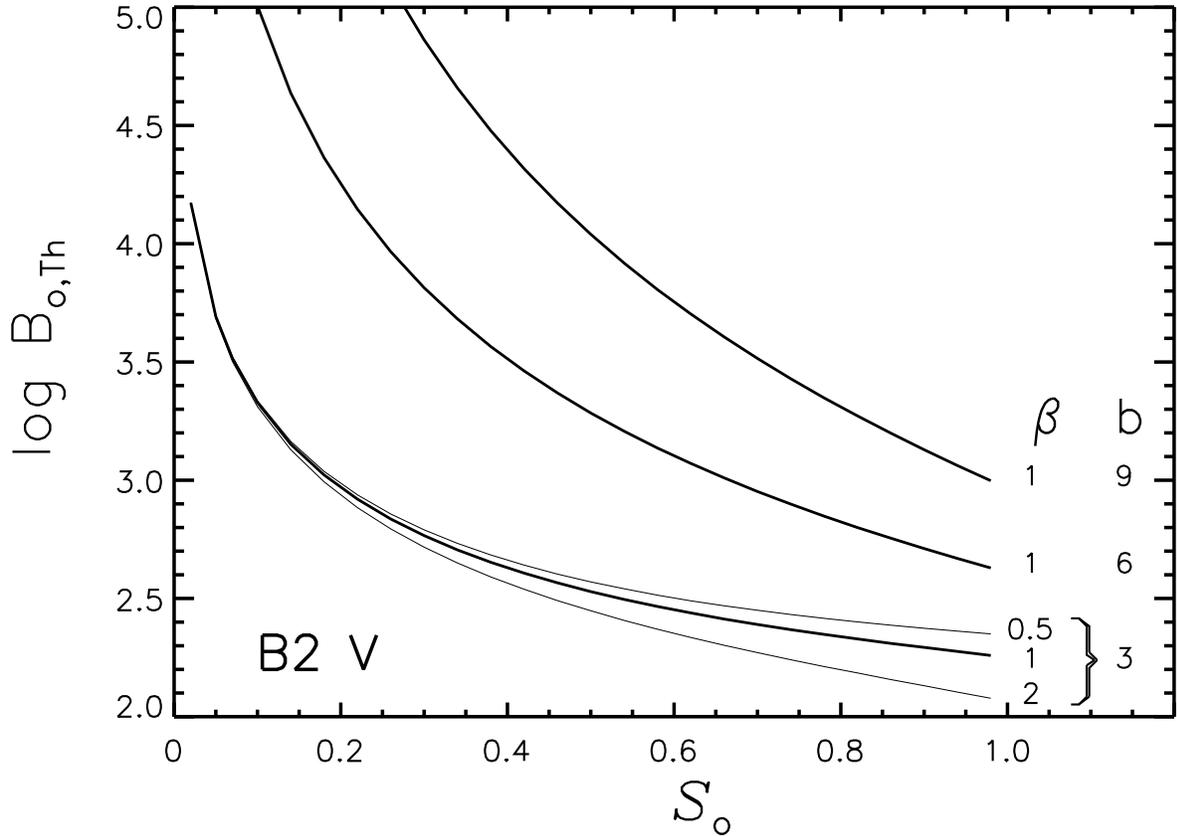}
\caption[]{The minimal or threshold surface magnetic field for the
formation of a disk versus \protect\SZ\, for which we have used the parameters
of a B2V star from Table 1. The different curves illustrate the
effects of varying the model parameters, $b$, which gives the radial
dependence of the magnetic field, and $\beta$, which is the outflow
velocity law parameter. For a case in which the field drops more
sharply with distance from the star the required surface field needs
to be larger. Three values of $\beta $ are shown for the $b=3$
case. The slower the rise in velocity, (e.g. $\beta=2$) the smaller
the field needed to produce the torqued disk, for a given \protect\SZ. }
\label{Fig:BTHSZ}
\end{figure}

Figure~\ref{Fig:BTHSZ} shows \BZTH\ versus \SZ\ for a B2~V star for
various values of $b$ and $\beta$. For the dipole case ($b=3$) the
threshold surface field for a star rotating at \SZ=0.7 is about 250
Gauss. If the field drops off more rapidly with distance from the
star, a significantly larger surface field is required to produce the
disk.  Primarily this is because the inner boundary, which is
identical to \xalfven\ for \BZTH, is not at the stellar surface but at
2.43 R, so for the larger values of $b$, a larger surface field is
needed to give the same field at \xalfven. The effect of changing the
wind velocity law index is shown by the curves labeled with various
values of $\beta$ in Figure~\ref{Fig:BTHSZ}, and these indicate that
the steepness of the wind velocity law is also of some importance in
determining the field strengths required to produce a disk.  For
example, in the case with \SZ=0.7, a slowly-accelerating flow with
$\beta=2$ can be channeled to the equator by a field only 70\% as
strong as that needed for the more-quickly-accelerating $\beta=0.5$
case.

\section{Properties of a Torqued Quasi-Keplerian Disk}

\subsection{Hot Main-Sequence Stars and the Possibilities for Disk Production}

To see how \BZTH\ depends on stellar parameters we note that $M$, $R$,
\Mdot, \vinf, along the main sequence are all essentially determined
by the spectral type \Teff\ as given in a table from Bjorkman \&
Cassinelli (1993).  Table 1 gives the factor $Y_*$ in
Equation~(\ref{XABdef}) which isolates the field dependence on stellar
properties.  Adopting $b=3$ and $\beta = 1$ we can then find
\BZTH(\SZ) as a function of \Teff\ from
Equation~(\ref{XABdef}). Table~1 also includes representative values
of \BZTH\ for \SZ =0.5, 0.7, 0.9.

Figure~\ref{Fig:mahesTHSZ} shows the comparison the minimal field from
Equation~(\ref{BZTHval}) along with the bounds on stellar magnetic
fields that were derived by Maheswaran \& Cassinelli (1988,
1992). The latter bounds were derived from considerations of envelope
circulation currents and maximal contributions of field pressure to
hydrostatic support.  We can see in Figure~\ref{Fig:mahesTHSZ} that
very large magnetic fields are required to produce disks around O
stars. Such large fields are needed because the large wind momenta
(\Mdot \vinf) of O-star require large magnetic forces to deflect their
flows.  For the case of the O3 V star the required field lies above the
maximal allowed field for many different stellar spin rates.  For the
early B stars with a spin rate of $\SZ=0.7$ the field required is in
the range 100 to 1000 Gauss. Such fields are small enough to have
avoided detection in the past, but are near the
field strength recently found by Donati \etal\ (2001) in $\beta$
Cep.  For the late B stars, we see that because of their small wind
momentum rates, rather weak dipole fields of order 30 Gauss could
produce a torqued disk.  In such cases we need to know if the 
predicted disks would be detectable.

The properties of the stars in Table 1 lead to specific predictions
regarding the disk extent and observational diagnostics should vary
with stellar parameters.  In particular, for given values of $b,
\beta, M, R$, we expect the radial extent of the quasi-Keplerian disk
to increase with increased \BZ\ while decreasing with increased
\Mdot\vinf, predictions that might be observationally testable. We now
investigate the radial extent of disks as function of stellar magnetic
field, and find the number of particles in the disk to find if the
model can explain the polarimetric properties and disk emission measures, 
which determine the \Halpha\ fluxes, of Be stars.

\subsection{Radial Extent}

In Section 3 we obtained expressions in terms of $\alpha$, \SZ, for
the radial extent of the disk, $\xinner \rightarrow \xouter$. This
extent is of course zero for $\BZ \leq \BZTH$. For $\BZ > \BZTH $, we
can obtain explicit values for \xinner\ and \xouter, in terms of
\BZ, \SZ\ and other stellar parameters (M, R, \Mdot, \vinf, T), by
translation from $\alpha$ to $\Gamma$ using
(\ref{gamdef}). Specifically, for any given set of stellar parameters
\Ystar, we can calculate $\Gamma $ and then obtain the corresponding $\alpha $
(for given $\beta, b $) from Figure \ref{Fig:gammaalpha}, or by
numerical inversion of (\ref{gamdef}). Then we have $ \xouter= \xalfven =
\sqrt{3}\alpha$. Here we illustrate results by adopting $b=3$, $\beta
=1$, and deriving \xinner\ and \xouter\ as functions of \SZ\ and the
ratio $\BZ/\Ystar$

The results are shown in Figure~\ref{Fig:XradvsB}.
Figure~\ref{Fig:XradvsB}a shows that results for three values of \SZ.
On this panel, it is easy to see that \SZ\ sets the threshold for the
field needed to produce a disk. This figure also shows that for values
of $B/\Ystar$ slightly above the threshold field, the disk extent increases
sharply, and in a seemingly homologous way vs. \SZ. To check the
apparent homology we show in Figure~\ref{Fig:XradvsB}b the ratio of
\xinner\ and \xouter\ expressed in terms of threshold radii \xTH\
versus the ratio of $B/\BZTH$.

As for the inner boundary, the position \xinner\ denotes the 
position in the disk where the
material first reaches the Keplerian speed (i.e. $\SK = 1$).  
For the minimum
surface field which will accelerate material to Keplerian speed
(\BZTH), the disk material reaches Keplerian speed at only one point, 
$\xinner=\xouter=\sqrt{3} \alpha$.  The other point on the 
kinematic \vphi\ curve in Figure \ref{Fig:vphivsx}
where the speed is Keplerian is of no interest because it
is beyond the torqued disk outer boundary, where the field is 
no longer dominant in determining the disk's $\vphi(x)$ structure.

Note in Figure~\ref{Fig:XradvsB} that, for values of \BZ\ large enough
to enforce nearly solid body rotation, the location of the inner
boundary is nearly independent of the magnetic field, as it is
primarily set by the value of \SZ\ as mentioned earlier.  In regards
to the outer boundary, an expression for $\BZ/\BZTH$ can be derived
from the ratio of two values of $\Gamma^2$ in Equation~(\ref{Gamdef}),
evaluated at $\alfa=\xalfven/\sqrt{3}$ and at $\alfa = \xTH/\sqrt{3}
$, and we find that the outer radius varies with $\BZ/\BZTH$, roughly
as $\xalfven/\xTH \sim (\BZ/\BZTH)^{2/5}$. The net effect is that the
outer boundary and the extent of the disk increases as \BZ\ increases,
but the location of the inner boundary does not change much.

From an observational point of view these considerations only
concern the magnetically torqued regions while the flow
in the far regions could also show enhanced densities because
of the WCD  effects as schematically illustrated in Figure 1.

The typical range of the disk model boundaries here is of the same
order with $\xouter \sim 3$ to 10, as discussed in the empirical
models of Be stars assuming Keplerian disks and based on IRAS
observations (Waters 1986), and \Halpha\ line observations by
Cot\'{e}, Waters, \& Marlborough (1996). As the latter authors note,
the observations do not rule out disks of greater radial extent, since
the tenuous outer parts contribute little to the polarization or
emission measure. Indeed some observations such as the radio
observations of Taylor \etal\ (1990) show that high density material
can reach distances of order 100 $R$.  The bulk of this extent might
be a tenuous outflow WCD/WCZ region as opposed to a quasi-Keplerian
one.

Now we have the physical boundaries of a Magnetically Torqued Disk and
expressions for the density distribution in the disk so it is possible
to test if the predicted disk satisfies basic observational limits,
especially in regards to the classic Be star diagnostics provided by
polarization and \Halpha\ observations.

\section{Comparisons with Observations}

\subsection{Continuum Scattering Polarization}

One important diagnostic of the Be disk properties is the polarization
they produce by electron scattering. Using the Brown \& McLean (1977)
general theory of (single) scattering polarization, one finds that for
a disk of approximately uniform thickness $H$ and a density
distribution given by Equation~(\ref{xabigequ}) for $\xinner \leq x
\leq \xouter$, the resulting polarization is

\begin{equation}
\label{polzatJC}
p=  \frac{3}{32}\frac{\sigma_T}{\pi}  \int_{\xinner}^{\xouter}
n_e(x) {\cal D}(x) H dx/x
\end{equation}
Here $\sigma_T$ is the Thomson cross section, $n_e$ is the electron number
density $(=\rhuD/\mu m_H)$, and we ignore absorption and assume the disk is
fully ionized.
Here we have included the finite source depolarization
factor, ${\cal D}$ of Cassinelli, Nordsieck, \& Murison (1987),
and Brown, Carlaw, \& Cassinelli (1991), and assumed
an edge-on aspect angle. Equation~(\ref{polzatJC}) yields, for $n_e(x)$
given by Equation (\ref{rhoin})

\begin{equation}\label{polzat1}
p=\frac{3 \sigma_T H}{32\pi} \NDZ I_p
\end{equation}
where the integral
\begin{equation}
I_p=\int_{1/\xouter}^{1/\xinner} y^{b-1}(1-y)^{\beta+1/2} dy
\end{equation}
can be expressed in terms of incomplete Beta functions.

For $\xouter \gg 1$, $\beta=0.5, $ and $ b=3$, the integral $I_p\approx 1/(2
\xinner^2) -1/(3 \xouter^3) \approx 0.1$ for typical $\xinner$.  Using
Equation~(\ref{rhoin}) for the disk density in
Equation~(\ref{polzat1}), the resulting numerical value for the
polarization is given by:

\begin{equation}\label{upperp}
p(\%) \simeq\ 0.5 \left(\frac{\NDZ}{10^{11}} \right)
R_{12} \times \frac{H}{R}
\end{equation}
Using the theoretical disk number density from Equation~(\ref{rhoin})

\begin{equation}
n(x)= \NDZ~ x^{-b}(1-1/x)^{\beta}= 1.8\times 10^{11} {\rm cm}^{-3}
\frac{\Mdot_{-9} v_8~ x^{-b}(1-1/x)^{\beta}}{R_{12}^2 T_4},
\end{equation}
then Equation~(\ref{upperp}) produces $p \simeq 1 \%$ for  the scale
values $\Mdot_{-9}, R_{12}, v_8, T_4$ all $=1$. Polarizations of
about 1\% are typical for the intrinsic polarizations of 
Be stars (McLean \& Brown 1978).

\subsection{Disk Emission Measure and Luminosity}

Key signatures of the presence of dense disks include the emission in Balmer
lines and the IR continuum excess over that from the stellar photosphere.
The relation between these emission properties, the disk density structure,
and the continuum polarization has been discussed by various authors (e.g.
Waters \& Marlborough 1992; Bjorkman \& Cassinelli 1990, Yudin 2001, Wood,
Bjorkman, \& Bjorkman 1997). Here, as another check of our derived density
distribution, we estimate the line and continuum emission that is 
expected. Both the Balmer lines and the IR continuum
emission arise collisionally, so at moderate optical depths their
luminosities $L$ take the form $L = 4\pi j \times EM$, where $EM = \int_V
n^2 dV$ is the emission measure of source volume $V$, and $j$ is the
appropriate emission coefficient in [ergs cm$^3$ s$^{-1}$] for each process,
which we assume to be constant through the source volume $V$ since the
cooled post-shock disk is taken to be isothermal.  We assume the disk has a
roughly constant thickness $H$ at all distances from \xinner\ to \xouter,
and use the density distribution in Equation~(\ref{rhoDval}) for
$\rhuD(x)$, to find

\begin{eqnarray}
EM &=& 2\pi R^2 H \int_{\xinner}^{\xouter}  n_{\rm D}^2 x dx \\
   &=& 2\pi R^2 H I_L \NDZ^2 \\
   &=& I_L \times EM_{\circ}
\label{EMdef}
\end{eqnarray}
where $I_L$ is given by

\begin{equation}
I_L(b,\beta, \xinner, \xouter)=\int_{\xinner}^{\xouter}
 x^{-2b+1} \left( 1-\frac{1}{x} \right)^{2\beta} dx
\end {equation}
The integral $I_L$ can be expressed in terms of incomplete beta functions.
For integer $b$ and half integer  $\beta$ it can be evaluated analytically
and we find it insensitive to $\beta$ and to $\xalfven \gg 1$. For $b=3$
its value is in the range $I_L= (1 - 5) \times 10^{-2}$ for 
$2 > \xinner > 1$.  The scale emission measure $EM_o$ is

\begin{eqnarray}
EM_{\circ} & = & 2 \pi R^2 H \NDZ^2 \\
     & = &   \frac{1}{8 \pi} \frac{H}{R}
             \left[ \frac{\Mdot \vinf}{R^{1/2} c_S^2 m_H} \right]^2\\
     & = & 2.3  \times 10^{61} \units{cm}^{-3}
         \frac{H}{R} \left[ \frac{\Mdot_{-9} v_8}{T_4 R_{12}^{1/2}}
	 \right]^2
\end{eqnarray}
If we use an approximate value for the emissivity,
$ \jHalp4pi = 2.4 \times 10^{-25}$ergs cm$^{3}$ s$^{-1}$ (Osterbrock 1989)
we find the line luminosity, $L(\Halpha)= \jHalp4pi \times EM$ to be

\begin{equation}
\label{Lhalpha}
L(\Halpha) = 6.3 \times 10^{36} {\rm erg~s}^{-1} \frac{H}{R}
\left[\frac{\Mdot_{-9}^2 v_8^2}{R_{12} T_4^2}\right] I_L
\end{equation}

 Also of interest is the comparison of the line luminosity $L(\Halpha)$
with the stellar continuum $L_\lambda$ at the \Halpha\ line wavelength. For
the purposes here
the stellar continuum luminosity can be approximated by the
Rayleigh-Jeans distribution $B^{RJ}_\lambda(T)$. Thus the monochromatic
stellar continuum luminosity at \Halpha\ is

\begin{equation}
\label{Llamstar}
L^{*}_{\lambda} \approx \frac{8 \pi^2 R^2 c k T}{ \lambda^{4}}
\end {equation}
This yields the optically thin line equivalent width
$W_\lambda =L(\Halpha)/L^{*}_{\lambda}(T)$
using Equations (\ref{Llamstar}) and (\ref{Lhalpha})  to be

\begin{equation}
W_\lambda = \frac{I_L}{64 \pi^3} \frac{\jHalp4pi \lambda^4}{c} \frac{H}{R}
  \frac{\Mdot^2 \vinf^2}{R^3 (kT)^3}
\end {equation}
giving, for $I_L \approx 0.05$

\begin{equation}
\label{Wlambda}
W_\lambda   =  1.3~ {\rm \AA}~  \frac{H}{R} 
\left( \frac{\Mdot_{-9}^2 v_8^2}{R_{12}^3 T_4^3} \right)
\end{equation}
For larger $\Mdot$ stars this yields a huge value for $W_\lambda $ since we have
ignored attenuation. We roughly correct for this by evaluating

\begin{equation}
W_\lambda '=W_\lambda  e^{-\tau}
\end {equation}
and use an approximate disk electron scattering optical depth

\begin{equation}
\tau \approx n_{\circ} R \sigma_T
\approx 0.1 \frac{\Mdot_{-9} v_8}{R_{12} T_4}
\end{equation}

Results for $\tau$, $W_\lambda '$ versus spectral type \Teff\ are
given in Table 2 and shown in Figure~\ref{Fig:HalEW}. For stars with
higher \Teff\ the high disk optical depth can suppress $W_\lambda '$
while for the lower \Teff the low density results in small $W_\lambda$
even before optical depth effects are taken into account. According to
this model, only a narrow range of \Teff\ around that of early B stars
should show an easily detectable \Halpha\ line.  Observationally,
Jaschek \& Jaschek (1983) found that the frequency of Be stars as a
function of spectral type peaks at spectral class B2, and a similar
result is seen in the IRAS survey of B stars by Cot\'{e} \& Waters
(1997). Equivalent widths of \Halpha\ range from a few \AA\ to about
30 \AA\ (Cot\'{e} \etal\ 1996; Cot\'{e} \& Waters 1997).  The
Magnetically Torqued Disk model predicts roughly the right emission
line properties, although clearly a better treatment of the radiation
transfer is required. Also the Be phenomenon extends to late B stars,
so a reason for the wide range of spectral classes that contain Be
stars needs to be explored in the context of our model. We now
consider the implications of this model for the qualitatively
different phenomenon, the variability of the Be stars on a variety of
time scales.

\subsection{Time Variable Phenomena}

Undoubtedly many of the chief clues to the dynamical structure of Be star
disks come from the complex time variability of these objects.  Be stars
vary on a wide range of characteristic times from ``activity'' on scales of
hours (Balona 2000) to quasi-periodic variability on years to decades
(Doazan 1982). Also of interest is the phase transition time from
non-emission line to emission line phases that indicate disks can vanish and
re-appear on time intervals of hundreds of days, (Hubert-Delplace \etal\ 1982
and Hubert 1981; Dachs 1987).  The connection between our Magnetically
Torqued Disk (MTD) model and these time scales is discussed below in
\S~\ref{sec:strength}. In addition to changes in overall line strength,
another interesting source of variability in these objects is the cyclic
change in the strengths of the violet and red peaks of the \Halpha\ emission
lines, known as V/R variation. This phenomenon comprises a shifting
asymmetry of emission line peaks from $V$ to $R$ over a timescale of order a
decade (Telting \etal\ 1994; Telting 2000). Ideas for understanding the $V/R$
variations within the context of the MTD model are discussed in
\S~\ref{sec:VR}

\subsubsection{Time Development of Disk Thickness}
\label{sec:strength}

Our interest has been in finding the field strength required to have matter
transmitted from the star to the relatively dense disk inferred from
observations. Later we mention some ideas for modifying our basic model to
produce a gas dominated disk that seems to be required for the one armed
instability to operate.

As can be seen in Equation (\ref{Wlambda}), we can attribute time
variability of \Halpha, to the time dependent increase in the height of the
disk, $H(t)$, so we need an expression relating the time derivative $dH/dt$
to stellar parameters. We first note that the mass inside a disk area
element $A(d)$ at any time is $\rho(d) A(d) H$. Both $\rho(d)$ and $A(d)$
should not change appreciably in our rigid field model. If we equate the
change in this mass to the mass influx from the magnetically channeled wind,
we obtain

\begin{equation}
\rho A(d) \frac{d H}{dt}= \FM(d) A(d)
\end{equation}
where $\rho(d)= \FM(d) \vw(d)/c_S^2$ 
and we get 

\begin{equation}
\frac{dH}{dt} = \frac{c_S^2}{v_w} = 
0.52 \times 10^{12} \units{cm/year}~ \frac{T_4}{v_8}
\end{equation}

Let us consider the time for the star to change from the non emission phase
to an emission phase Be star. For this let us assume the lines
are noticeable when the equivalent width is about 1 \AA. Therefore, setting 
$W_{\lambda}= 1.0$ in Equation (\ref{Wlambda}), we obtain 
from $H(t)  = (dH/dt) \times t_1$, the start-up time,

\begin{equation}
\label{timeone}
t_1= 1.5 \units{years}~ \frac{R_{12}^4 T_4^2}{\Mdot_{-9} v_8}
\end{equation} 
For the values associated with a B2 star, and with $T_4 = 0.8 \Teff$ this
gives a reasonable value of about 4 months. For later type B stars, the value
increases significantly (100 years in the case of a B5V stars and $10^3$
years at B9V). These long times for the late B stars is mostly because the
mass loss rates we assumed for these stars are much lower, and hence $t_1$
in Equation (\ref{timeone}) is longer. Although there is a known a tendency
for later B stars to change on a slower time scale (Hubert-Delplace \etal\
1981), we suggest that a better agreement of our model with the observations
of the late B stars would occur if we accounted for the extreme version of
FMR theory that was called ``Centrifugal Magnetic Rotator'' theory (L \& C,
\S 9.7.1 and Figure 9.7). In this class of magnetic rotator winds the mass
loss rate is enhanced because the density scale height in the subsonic
and solid body portion of the wind is increased by centrifugal forces and
\Mdot\ increases exponentially with $\SZ^2$ (L \& C, Eq 9.90). The late Be
stars appear to have a larger \SZ\ on average than do the early Be stars
(Hubert-Delplace \etal\ 1981), and would be the ones for which centrifugal
magnetic rotator effects are likely to be important. If late Be stars have
mass loss rates in excess of those predicted by pure line driving, our
estimate for the disk start-up timescale could be brought to an reasonable
range.

In our model, the height of the disk would increase linearly with time,
so a growth-time scale can be defined as $ t_G = 1/ (d\ln H/dt)$. 
Of particular interest in this regard is the fill-up time $t_{fill}$.
This is the time it takes the incident mass flux to increase the
height $H$ of disk matter in the tube to its maximal value which is about
$d$ (c.f. Donati \etal\ 2001 discussion). Thus

\begin{eqnarray}
t_{fill} &\simeq& \frac{d}{\frac{dH}{dt}} =
\frac{d}{v_w}\left(\frac{v_w}{c_S}\right)^2
= t_{Flow} \left(\frac{v_w}{c_S}\right)^2 x\\
        &\simeq & 2.1  \units{years} ~~ \frac{R_{12} v_8}{T_4} 
	\times x
\end{eqnarray}
Made without considering detailed mechanisms, this estimate of the
disk filling time is much longer than the wind flow time ($t_{Flow}
= R/v_\infty$).  The disk filling time is  about four years in the
inner disk ($x \simeq 2$), with the outer regions eventually filling
in around a decade---tantalizingly close to the observed long term Be
star disk variability time scales. Some discussion of specific field
disruption mechanisms (reconnection, diffusion) by matter accumulation
is to be found in Havnes \& Goertz (1984), Babel \& Montmerle
(1997a), and Donati \etal\ (2001).

\subsubsection{Connections with the V/R Variations}
\label{sec:VR}

The interpretation of the $V/R$ variations proposed by Papaloizou \etal\
(1992) and Okazaki (1991) is that the variation is due to the very slow
rotation/precession of an azimuthally localized density enhancement in the
disk across the line of sight from the approaching to the receding part of
the disk. The physical model proposed for such a density enhancement and
drift is that it is a density wave in the disk, somewhat like galactic
spiral density waves, but attributable to the gravitational field of the
non-spherical central star, rather than to self-gravitation (negligible in a
stellar disk). This model is attractive in predicting a density spiral
precession period of the right order for typical Keplerian speeds. The MTD
model discussed here shares some important features with the models of
precessing one-armed spirals.  Most importantly, both models envision a disk
in which material spends a long time in roughly Keplerian motion (in fact,
the observations of $V/R$ variations are often cited as evidence of the
existence of a near-Keplerian disk). 

The V/R variability on irregular time scales of average 7 years
(Okazaki 1997) has been explained as the orbital motion of a one armed
spiral pattern in a disk. A model by Lee, Saio, \& Osaki (1991) has
proposed that the matter is driven up from the star by viscosity. A
one armed pattern in the disk has been explained by the Global Disk
Oscillation model of Okazaki (1991, 1996) and Papaloizou \etal\
(1992). The GDO model would operate in quasi-Keplerian disks but not
in disks in which the material has a short residence time. Matter must
be supplied to the disk either continually or by events such as
non-radial pulsations that occur on a time scale shorter than the
viscous time scale as considered by Owocki \& Cranmer (2001).

The V/R phenomenon is one that all models must try to address. In our
magnetic channeling/torquing model, for strong fields a substantial part of
the disk would have corotating $v_\phi$ values and one might expect the
field torquing forces to overwhelm any gravitational density wave effect,
though in the outer disk where the field will bend (attaining a $\phi$
component) there might be some such effect.  Any concentration of disk
density by a stellar surface field that is not uniform in $\phi$ would be
expected roughly to corotate with the star, rather than over long periods.
This is therefore an intriguing open question for future consideration.

A very interesting idea, proposed by H. Henrichs (2002, private
communication), is that the stellar field might be time
variable. During periods when it is strong our model would apply and a
dense disk would build up over the fill time we estimated.  However as
it stands the model provides no obvious way to explain $V/R$
variations, at least via density wave precession in a Keplerian disk
controlled by the gravity of the non-spherical star. If, however, the
field weakened substantially then the disk would decouple from the
field and become near Keplerian through viscous effects and capable of
supporting such density waves until the field returned or the disk
decayed - another observed phenomenon the timescale of which requires
explanation. This idea makes the testable prediction that $V/R$
variations should be absent during disk build up and present only at
times of weak field.

\section{Discussion}

\begin{figure}
\plotone{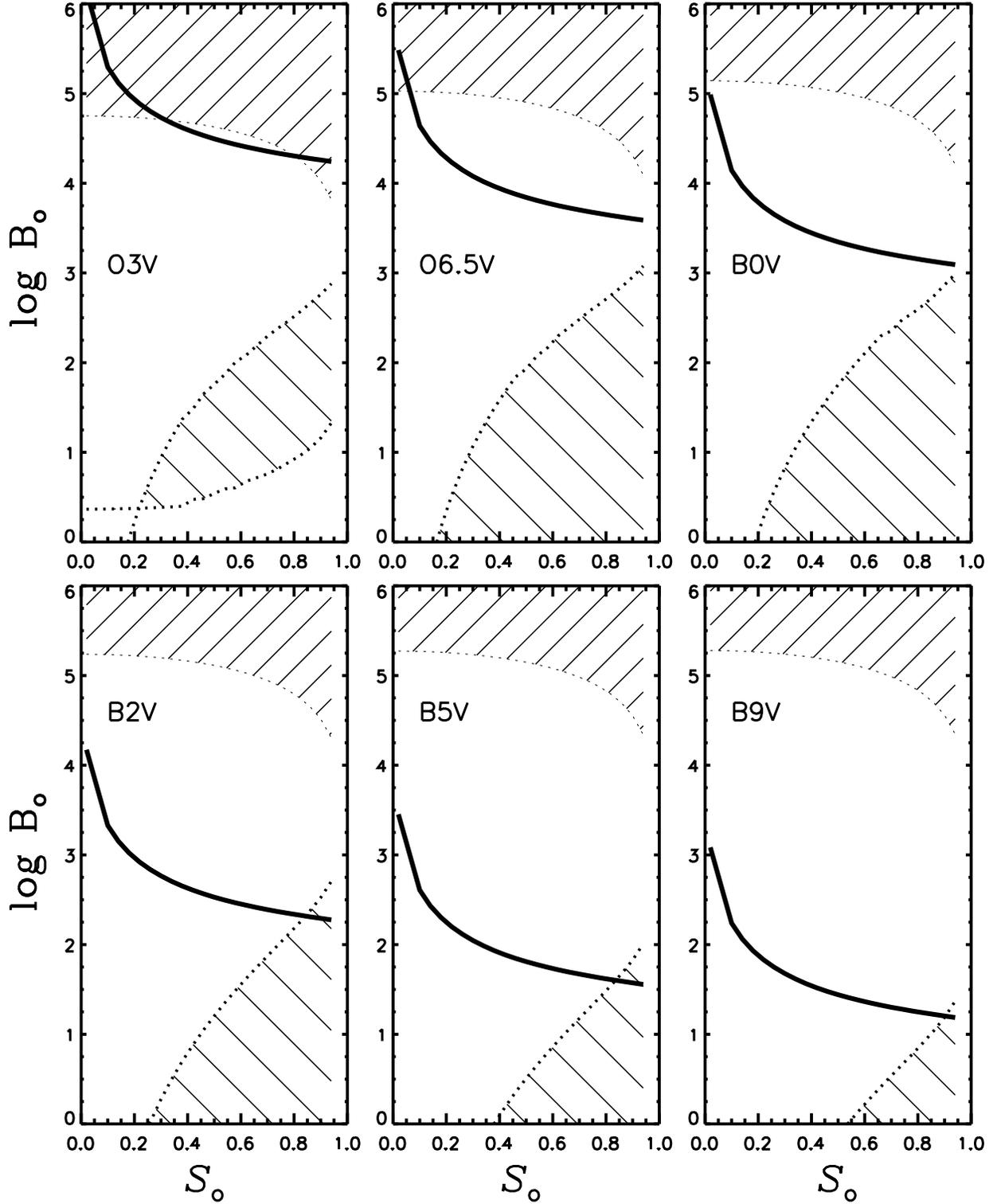}
\caption[]{Shows the limits derived for the maximal and minimal field
strengths for Be stars versus \protect\SZ\ derived from photospheric
constraints of Maheswaran \& Cassinelli (1992). The solid line shows
the minimal \protect\BZ\ as a function of \protect\SZ\ such that a disk can be
magnetically torqued; for these calculations we use the stellar parameters
from Table 1, $b=3$, and $\beta = 1$.}
\label{Fig:mahesTHSZ}
\end{figure}

\begin{figure}
\epsscale{1.0}
\plotone{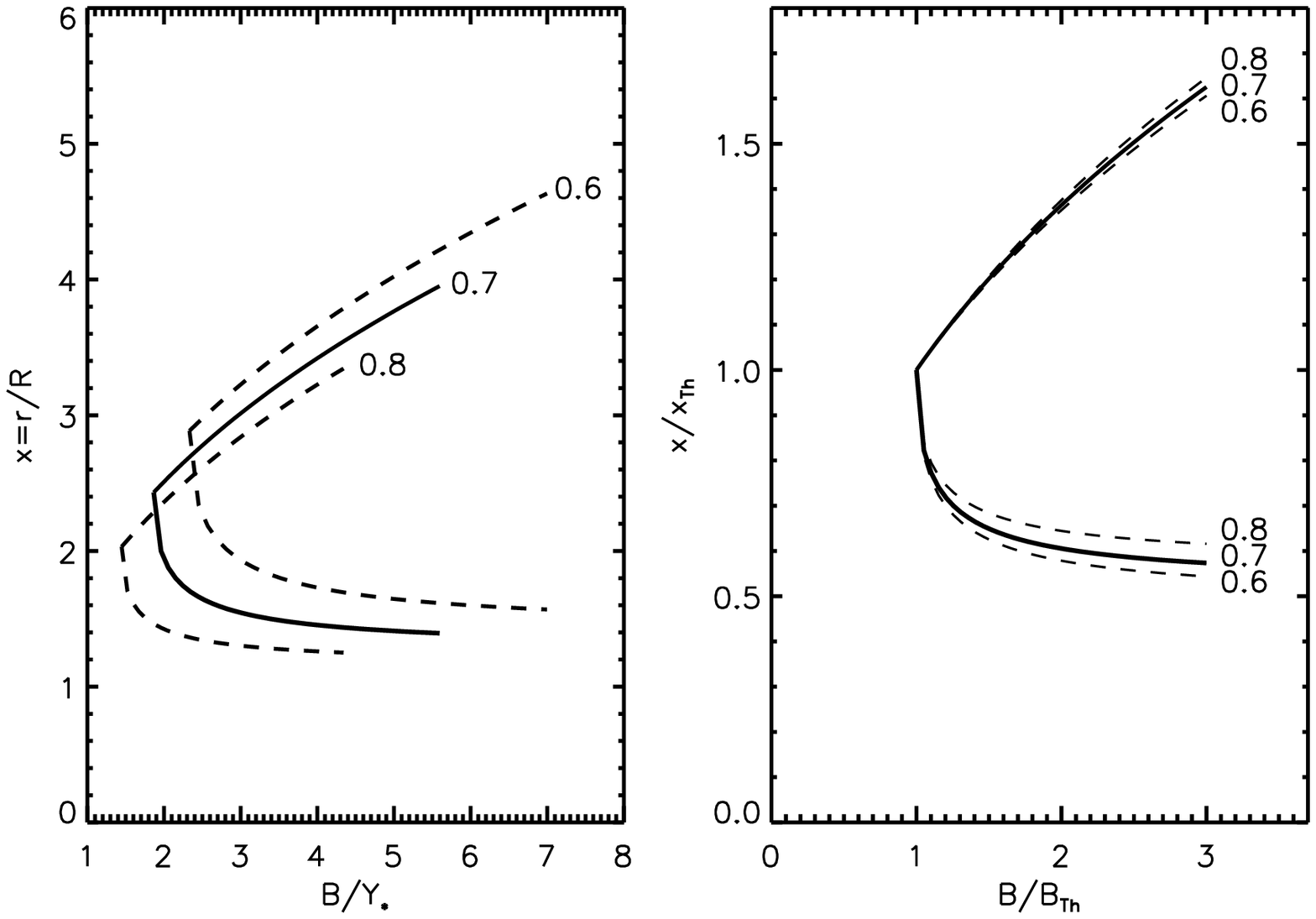}
\caption[]{Two panels that show the radial extent (\protect\xinner\ to \protect\xouter) 
of magnetically torqued disks versus the surface magnetic field, \protect\BZ. 
The $B$ field is normalized in units of \protect\Ystar, so the plot is applicable 
to any star. a)  In the left panel
results for the boundaries are shown for three values of the spin rate
parameter \protect\SZ. The upper segments of the curves correspond to \protect\xouter\ and
note these merge with the lower segments, the \protect\xinner\ curves, at the 
threshold point for each of the three models. The curves appear very
similar in shape although shifted.  b) To illustrate the homologous nature
of the curves, in the right panel is shown the quantities on both axes in 
the left panel are normalized by the threshold values. 
Notice that the extent of the disk (\protect\xouter - \protect\xinner) increases sharply
for magnetic fields just above the threshold value.}
\label{Fig:XradvsB}
\end{figure}

The central point of our model is that magnetic torquing and channeling of
wind flow from the intermediate latitudes on a B star can, for plausible
field strengths, create a dense disk a few stellar radii in extent in which
the velocity is azimuthal and of order the local Keplerian speed. The model
was motivated by the fact that line profile data (e.g. Hanuschik 1996)
indicate that disk emission lines of Fe II, extend to $\Delta v > v_{\circ}
\sin i$, and reach disk Keplerian values. Also the sharp 
absorption features seen in the center of those lines do not show evidence
for inflow or outflow. The latter of course would be expected in the
case of the WCD model. The present paper aimed to show that fields could confine and torque
the flow to create a quasi-Keplerian disk of roughly the spatial extent
suggested by these observations. Future work to test the real viability of
the model will address two important questions:

\begin{enumerate}   

\item[a)] Is the azimuthal velocity distribution $\vphi(d)$ across
the radial range of the dense disk predicted by our magnetic model
compatible with observed line profiles? This model distribution starts
as roughly corotational, \ie\ $\vphi(d) \propto d$, and transitions
to a \vphi\ decreasing with $d$. This distribution is clearly not truly
Keplerian disk $v_\phi(d) \propto 1/ \sqrt{d}$. However, the typical
$v_\phi$ is comparable with $v_K$ and exists over only a rather narrow
range of $d$, unless the field is very strong, which is unlikely on
theoretical grounds (Maheswaran \& Cassinelli 1988, 1992;
c.f.~Figure~\ref{Fig:mahesTHSZ}). Outside the model disk domain, \ie\
at $x> \xalfven$, we predict quantitatively a somewhat flattened WCD
region changing with increasing $d$ to a WCZ.  At even farther
distances from the star, the wind will be nearly spherical.  All of
these regions involve strong outflow velocities but with densities
much below those in the quasi-Keplerian disk. It is important to
estimate whether the high $v_r$ but low density distributions there
are compatible with line profile data \ie\ does there exist an outer
expanding flattened disk which affects line profiles very little but
is detectable by radio observations etc. These and other diagnostic
predictions of the Magnetically Torqued Disk model such as X-ray
production will be treated in subsequent papers.

\item[d)] Another issue, that is perhaps related to the $V/R$ variations
is the winding up of the field - something we have ignored so far. In a model
with strictly corotational $\vphi(r)=\vZ r/R$ out to some point and angular
momentum conservation beyond there, the field would corotate in the disk
region. However, in practice and for our parametric model (Equation
\ref{vphi/v0}), \vphi$(r)$ falls behind corotation and the field will be
wound up. The  resulting \Bphi\  component will have opposite sign on opposite
sides of the disk plane and will grow until field reconnection can occur
allowing a quasi-steady state with the disk matter slipping across the field
lines. In such a situation gravitational control of the disk may dominate
sufficiently for the disk to behave as quasi-Keplerian and allow density
waves. Indeed, in the light of this, one might argue that our estimate of
$\BZTH$ is too conservative since the crucial feature of our model is that
the field should deliver wind matter to the disk with sufficient angular
momentum (from magnetic torquing) and with essentially no $v_r$ but only
$v_\theta$ so that shock compression can occur without any outflow (in
contrast to the WCD model). This needs working out in detail but if only the
wind density needs torquing action our field limits would be much reduced.
Furthermore, in that case the disk $\rho_D v_\phi^2$ would be so high
compared to $B^2/8\pi$ that the matter would completely dominate the field
once delivered into the dense disk, which would consequently be essentially
Keplerian and so support density waves under the action of stellar gravity.

\end{enumerate}

In this paper we have followed a used a kinematic description of the
azimuthal velocity of a disk zone, along with constraints based on magnetic
rotator theory to develop an analytic model which we call the Magnetically
Torqued Disk (MTD) model, We derived the independent parameters for disk
formation problem: the spin rate parameter. \SZ, the surface field \BZTH\
and the stellar parameters contained in the quantity, \Ystar. We assumed a
rather simple rigid magnetic field picture, but one that allowed us to
derive an estimate of the field required to produce a torqued disk for main
sequence stars ranging from O3 V to B9 V. Although an unreasonably large
field was found to be needed for the early O stars, we concluded that at the
early B spectral class at which the Be phenomena are most prominent, the
model provided about the right observational properties, and for reasonable
surface field strength. Also, we considered several time scales associated
with Be stars, and found start-up, growth and fill-up time scales that are in
the month to decade long range that seem appropriate. For the late B stars
the start-up time scale and the \Halpha\ equivalent widths appear to be small
and we suggest that perhaps the mass loss rates for these stars need to be
enhanced by centrifugal magnetic rotator forces. Modifications that are
needed to address V/R variations and observational consequences at
a variety of wavelengths will be considered in later papers.

\acknowledgements We gratefully acknowledge the financial support of a
PPARC Fellowship (JPC) and PPARC Grant (JCB), Studentship (DCT) and
NSF grant AST-0098597 (JPC). The paper has benefited from discussions
with J.~Mathis, R.~Hanuschik, L.~Oskinova, J.~Bjorkman, and
A.~Lazarian.

\clearpage

\begin{table}
\setlength{\tabcolsep}{1mm}
\begin{center}
\caption{Minimum fields for torqued disks around stars along the main
sequence }
\begin{tabular}{lccccccccc}
\tableline
\tableline
 & & & & & & & \multicolumn{3}{c}{\BZTH}\\ \cline{8-10} 
Spectral & $\Teff$  & $R$ & $M$ &  \Mdot  & \vinf & $Y_*$
& \SZ=0.5 & \SZ=0.7 &
\SZ=0.9\\
 Type   & ($10^4$ K)    & ($10^{12}$ {\rm cm})& (10 \Msun) & ($10^{-9} \Msunyr$)  & (\kms)
 & (G)  & (G) & (G) & (G) \\
\tableline
O3   & 5.0& .98 & 5.5 & 9100   &3100 & 12700 &30900  & 22400 & 17900 \\
O6.5 & 4.0& .70 & 2.9 &  310   &2500 & 2830  &6890  & 5000  & 4000 \\
B0   & 3.2& .46 & 1.5 &   27   &1300 & 904   &2200  & 1600  & 1280\\
B2   & 2.3& .31 & .83 &  .4    & 840 &  138  &335  & 243 & 195 \\
B5   & 1.5& .23 & .45 &  .01   &  580 &  26 & 64 &  46 & 37 \\
B9   & 1.0& .18 & .26 &  .0013 & 460  & 11  & 27 &  20 & 16 \\
\tableline
\end{tabular}
\end{center}
\label{Tab:startable}
\end{table}

\clearpage

\begin{table}
\vspace{0.5cm}
\begin{center}
\caption{Equivalent widths of $H_\alpha$ from torqued disks in stars along
the main sequence}
\vspace{0.1cm}
\begin{tabular}{lcrlcl}
\tableline
\tableline
Spectral & $\Teff$  & $L(\Halpha)$  & $W_\lambda$ & $\tau$ & $W_\lambda$~$e^{-\tau}$ \\
  Type   &($10^4$ K) & (ergs s$^{-1}$) & (\AA) &   & (\AA)    \\
 \tableline
O3   & 5.0& $3 \times 10^{40}$ & $2.0\times 10^6$ & 580   & 0.0     \\
O6.5 & 4.0& $1 \times 10^{39}$ & $8.2\times 10^3$ & 27 & $7.8\times 10^{-9}$   \\
B0   & 3.2& $9 \times 10^{37}$ & 114       & 2.4 & 10.65       \\
B2   & 2.3& $1 \times 10^{36}$ &  0.089    & .046 & 0.085       \\
B5   & 1.5& $4 \times 10^{34}$ & $2.4\times 10^{-4}$ & $2\times 10^{-3}$ &
$2.4\times 10^{-4}$ \\
B9   & 1.0& $8 \times 10^{33}$ &  $1.8\times 10^{-5}$ & $3\times 10^{-4}$ & 
$1.8\times 10^{-5}$ \\
\tableline
\end{tabular}
\end{center}
\label{Tab:halphaEW}
\end{table}

\clearpage

\clearpage

\clearpage

\clearpage

\clearpage

\clearpage

\clearpage

\clearpage

\clearpage

\clearpage

\begin{figure}
\epsscale{1.0}
\plotone{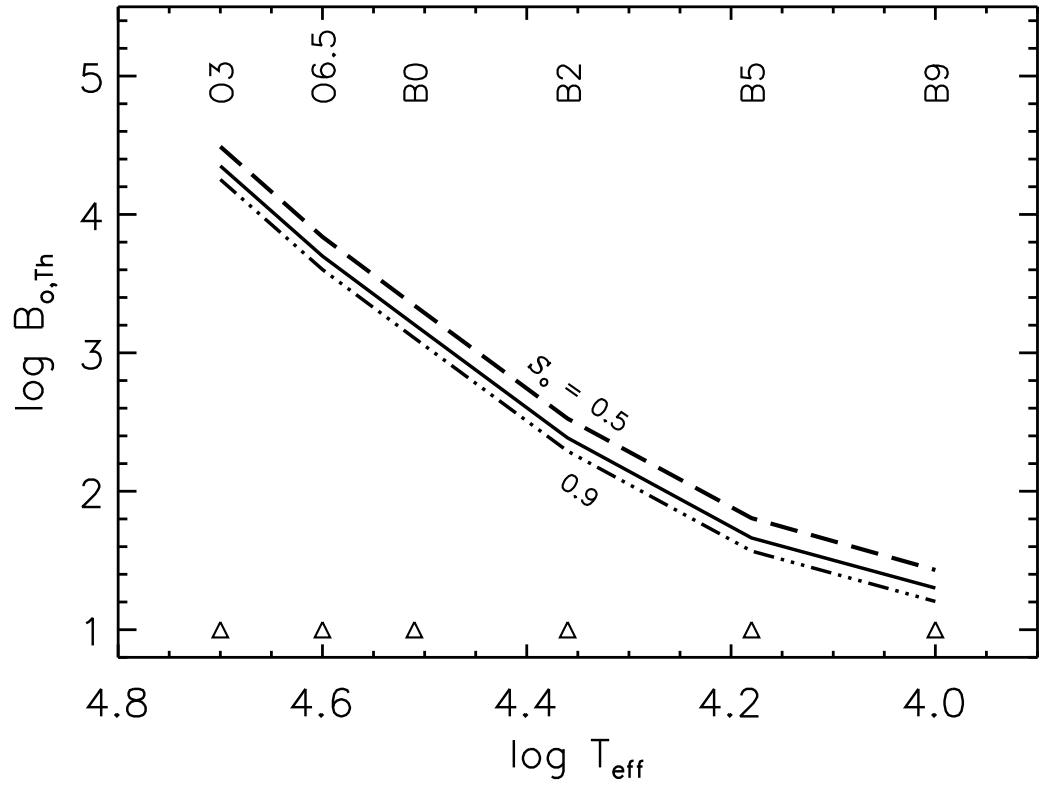}
\caption[]{Shows the threshold field for torqued disks for stars along the
main sequence. The stellar parameters are from Table 1 and $b=3$, $\beta =1$
are assumed. The three curves correspond to the values of
the spin rate \protect\SZ\ = 0.5, 0.7 and 0.9, two of which are labeled.}
\label{Fig:B_vsTeff}
\end{figure}

\clearpage

\begin{figure}
\epsscale{1.0}
\plotone{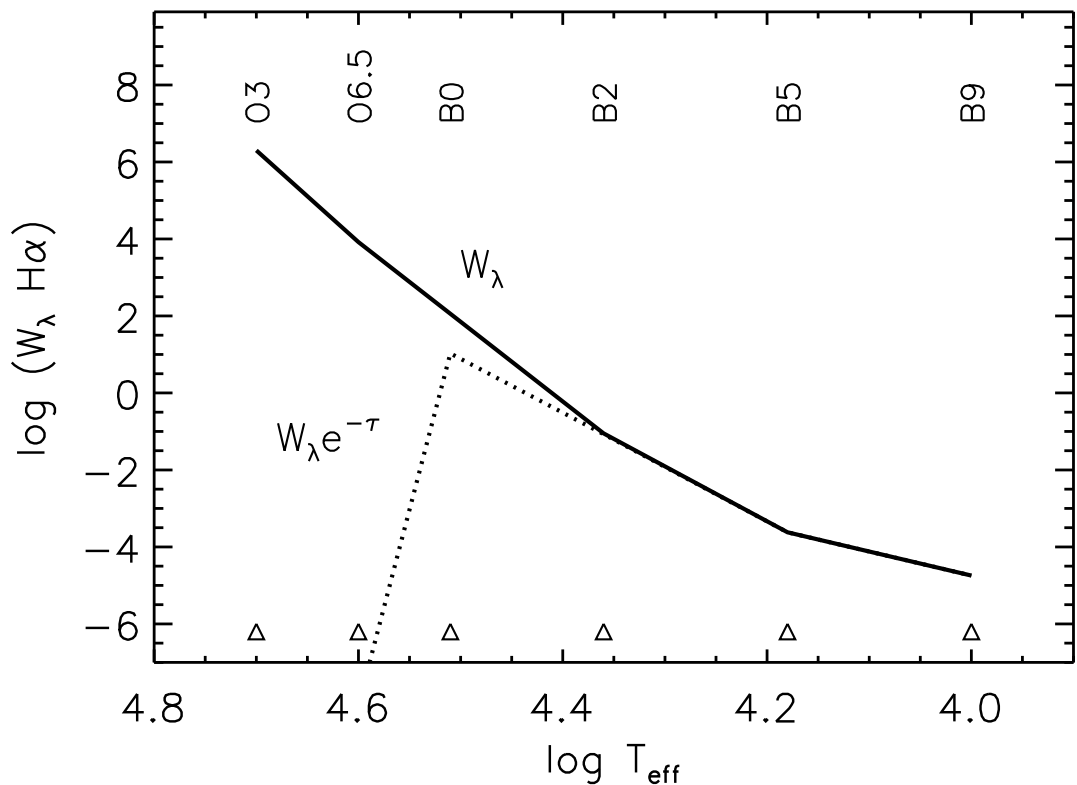}
\caption[]{The solid line shows an estimate of the upper limit to the
equivalent width of the \protect\Halpha\ line in \AA\ versus \protect\Teff, produced
in a disk of infinite extent. The dotted line shows the reduced value
of the equivalent width when optical depths in the disk are accounted
for in an approximate way.  The unmodified $W_\protect\lambda$ indicates
unacceptably large values for the early type stars, however, for these
stars the magnetic field required would also be excessively large
(i.e. greater than a kilogauss).}
\label{Fig:HalEW}
\end{figure}

\end{document}